%% file: main.tex
\PassOptionsToPackage{dvipsnames}{xcolor}
\documentclass[manuscript,screen, nonacm]{acmart}

\usepackage{wrapfig}
\usepackage{booktabs} 
\usepackage{graphicx} 
\usepackage{makecell} 
\usepackage{numprint} 
\usepackage{multirow} 
\usepackage{enumitem}
\usepackage{colortbl}
\usepackage{tablefootnote}
\usepackage{subcaption}
\usepackage{url}
\usepackage{longtable}
\usepackage{threeparttable}
\usepackage{hyperref}
\usepackage{tabularx}
\newcolumntype{Y}{>{\raggedright\arraybackslash}X}
\newcommand{\papersep}{\noalign{\vskip 3pt}}

\usepackage{tikz}
\usetikzlibrary{calc,positioning,arrows.meta}

\usepackage{bm}
\usepackage{etoolbox}
\usepackage{marginnote}

\makeatletter
\let\oldmarginnote\marginnote
\renewcommand*{\marginnote}[1]{%
    \begingroup%
    \ifodd\value{page}
        \if@firstcolumn\normalmarginpar\fi
    \else
        \if@firstcolumn\else\normalmarginpar\fi
    \fi
    \oldmarginnote{\textcolor{Brown}{#1}}%
    \oldmarginnote{}%
    \endgroup%
}
\makeatother

\AtBeginDocument{%
  }

\acmConference[Conference acronym 'XX]{Make sure to enter the correct
  conference title from your rights confirmation emai}{June 03--05,
  2018}{Woodstock, NY}

\begin{document}

\title{A Survey on Open Dataset Search in the LLM Era: Retrospectives and Perspectives}



\author{Pengyue Li}
\email{pengyueli@whu.edu.cn}
\orcid{0009-0007-2163-9566}
\affiliation{%
  \institution{School of Computer Science, Wuhan University}
  \city{Wuhan}
  \country{China}
}

\author{Sheng Wang}
\orcid{0000-0002-5461-4281}
\affiliation{%
 \institution{School of Computer Science, Wuhan University}
 \city{Wuhan}
 \country{China}}
 \email{swangcs@whu.edu.cn}

\author{Hua Dai}
\orcid{0000-0003-2465-8977}
\affiliation{%
  \institution{School of Computer Science, Nanjing University of Posts and Telecommunications}
  \city{Nanjing}
  \country{China}}
\email{daihua@njupt.edu.cn}

\author{Zhiyu Chen}
\orcid{0000-0002-3096-7912}
\affiliation{%
  \institution{Amazon.com, Inc.}
  \city{Seattle}
  \country{USA}}
\email{zhiyuche@amazon.com}

\author{Zhifeng Bao}
\orcid{0000-0003-2477-381X}
\affiliation{%
  \institution{School of Electrical Engineering and Computer Science, The University of Queensland}
  \city{Brisbane}
  \country{Australia}}
\email{zhifeng.bao@uq.edu.au}

\author{Brian D. Davison}
\orcid{0000-0002-9326-3648}
\affiliation{%
  \institution{Department of Computer Science and Engineering, Lehigh University}
  \city{Bethlehem}
  \country{USA}}
\email{davison@cse.lehigh.edu}

\renewcommand{\shortauthors}{Li et al.}

\begin{abstract}
High-quality datasets are essential for data-driven tasks such as training models for medical diagnosis, analyzing urban mobility, and supporting scientific research. However, the growing scale, heterogeneity, and semantic complexity of open data repositories, together with increasingly diverse dataset needs, make it challenging for users to efficiently identify datasets that best satisfy their requirements. Consequently, a systematic review of recent advances in open dataset search is both necessary and timely. {In this survey, we focus on developments beyond the conventional keyword--metadata matching paradigm. We first review the foundations of open dataset search, including dataset sharing and discovery infrastructures and representative search engines. We then organize recent techniques according to three major search paradigms: content-aware dataset search, dataset relationship search, and task-oriented dataset search. We further examine how large language models (LLMs) are reshaping the dataset search process, including search evidence construction, dataset representation and indexing, user input and search planning, and relevance estimation. We also review representative benchmarks and evaluation practices for dataset search.} Finally, we discuss open issues and promising future research directions. Through this structured review, we aim to clarify the evolving landscape of open dataset search and provide a foundation for future research in the LLM era.
\end{abstract}

\ccsdesc[500]{General and reference~Surveys and overviews}
\ccsdesc[500]{Information systems~Information retrieval}

\keywords{Open Dataset Search, LLM for Dataset Search}


\maketitle

\input{sections/sec1}
\input{sections/sec2}

\input{sections/sec3}

\input{sections/sec4}
\input{sections/sec5}
\input{sections/sec6}


\bibliographystyle{ACM-Reference-Format}
\bibliography{reference/final}



\end{document}

%% file: sections/sec1.tex
\begin{sloppypar}

\section{Introduction}\label{sec: Introduction}

In the era of big data and AI, datasets have become indispensable across a wide range of fields. From healthcare and finance to environmental monitoring and the social sciences, they form the foundation of many data-driven tasks. However, with the rapid growth of available data and the increasing sharing of datasets~\cite{benjelloun2020google}, a significant challenge has emerged: how to enable users to conveniently and accurately identify datasets that satisfy their data needs. This challenge, commonly referred to as the open dataset search problem, has attracted growing attention from both academia and industry~\cite{paton2023dataset,chapman2020dataset,fan2023table}.

\begin{figure*}[t]
\includegraphics[width=0.8\textwidth]{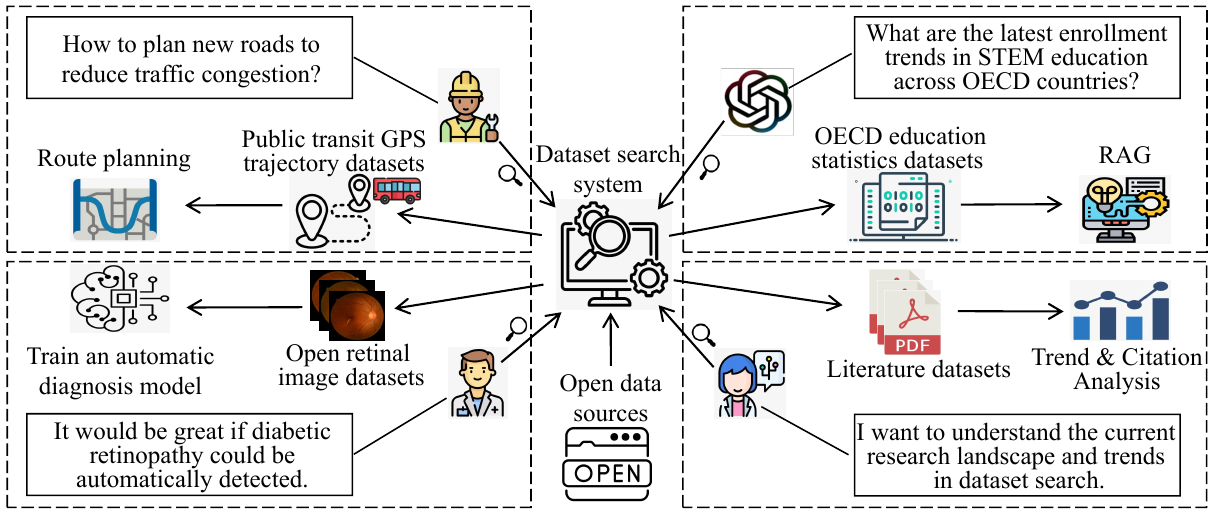}
\caption[]{Motivating scenarios for open dataset search.}
\label{fig: overview}
\end{figure*}

Open dataset search supports data discovery and reuse across a wide range of application domains. 
As shown in Figure~\ref{fig: overview}, in healthcare, retinal image datasets can be discovered to train models for automated diabetic retinopathy diagnosis. In urban planning, public-transit GPS trajectory datasets can support congestion analysis and route optimization. Researchers can search literature datasets to analyze research trends and citation structures in their fields of interest. For LLM-based applications, retrieved datasets can also provide external evidence for retrieval-augmented generation (RAG)~\cite{arslan2024survey}.
However, the growing scale, heterogeneity, and semantic complexity of available datasets, together with increasingly diverse and sophisticated search needs~\cite{grubenmann2018financing,hai2023data,kassen2013promising,yang2024budgeted}, pose significant challenges for open dataset search. A systematic review of existing techniques, remaining challenges, and promising future directions is therefore both necessary and timely.

\subsection{Open Dataset Search: Definitions}\label{subsec: Dataset and Dataset Search}

To establish a clearer understanding of the open dataset search problem, we first introduce formal definitions of the key concepts involved.

\begin{definition}[\textbf{Open Dataset}]\label{def: dataset}
An open dataset $D_i = \{o_{i,1}, o_{i,2}, \dots, o_{i,n}\}$ is a collection of related observations shared beyond its producer and made discoverable through a repository or platform accessible to a search system, where each $o_{i,j} \in D_i$ denotes a unit of dataset content. The dataset is accompanied by metadata, such as the title, description, schema, keywords, and license.
\end{definition}

\begin{definition}[\textbf{Open Dataset Search}]\label{def: Open Dataset Search}
Given a collection of open datasets $\mathcal{D} = \{D_1, D_2, \dots, D_m\}$ and a user input $Q$, open dataset search refers to the process of retrieving datasets from $\mathcal{D}$ according to their relevance to $Q$.
\end{definition}

\begin{figure*}[t]
\includegraphics[width=0.85\textwidth]{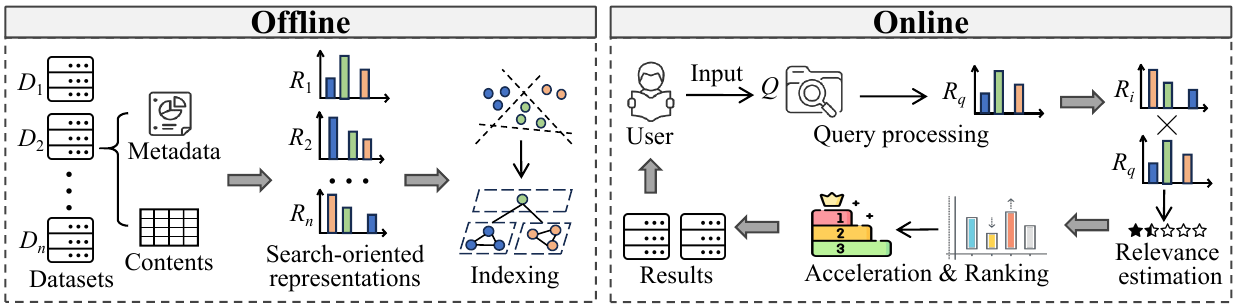}
\caption{A unified framework for open dataset search.}
\label{fig: example}
\end{figure*}

{
Here, \emph{open} emphasizes sharing and discoverability; the reviewed methods also apply to datasets shared within platforms or organizations, provided that the search system can access the required metadata or content. According to Definition~\ref{def: Open Dataset Search}, the core of dataset search is to assess the relevance between a user input and candidate datasets. Figure~\ref{fig: example} illustrates a general framework. Consider, for example, a researcher searching for public-transit GPS trajectory datasets for congestion analysis in a target city. Offline, dataset metadata and contents can be processed into search-oriented representations that preserve information useful for
relevance estimation. In this example, dataset titles and descriptions
may describe the covered city, transportation mode, and collection
period, while the dataset contents contain GPS trajectories and their
spatio-temporal distributions. The textual evidence can be encoded
into semantic vectors, while the trajectory contents can be summarized
using spatial or statistical representations. Indexes may then be
constructed over these representations to support efficient access.
Online, the user's input is processed into a form suitable for
relevance estimation against candidate datasets. For example, a
natural-language request such as ``find public-transit trajectory
datasets for congestion analysis in this city'' can be encoded into a
semantic representation and matched against dataset descriptions and
contents. Alternatively, if the user provides an exemplar trajectory
dataset, its spatial distribution can be represented and compared with
candidate datasets using a dataset-level distance. For task-oriented
search, the user input may additionally specify a downstream task, in
which case relevance can be estimated through task suitability or
downstream utility rather than direct similarity. Datasets are ranked
according to the resulting relevance scores, and the most relevant ones
are returned to the user. During search, indexes and acceleration
strategies, such as candidate filtering and pruning, help avoid
unnecessary data access and relevance computations, thereby enabling
efficient search over large dataset collections.}

\subsection{The Evolution of Open Dataset Search Techniques}

Many deployed dataset search engines and portals still rely heavily on keyword--metadata matching~\cite{brickley2019google,chapman2020dataset}, where the keywords provided by users are matched against dataset metadata. This paradigm has been widely adopted in real-world data portals due to its scalability and low infrastructure requirements. {However, modern open dataset search faces several key challenges: \textit{(1) dataset needs are becoming increasingly diverse and complex}~\cite{wei2026kats,lin2025datascout}, making them difficult to express and interpret through simple keyword queries. \textit{(2) dataset relevance is difficult to characterize from metadata alone}~\cite{chen2023dense,chen2024dunks}, as relevance may depend on dataset contents, inter-dataset relationships, or task-specific suitability and utility. \textit{(3) dataset search must accommodate heterogeneous data modalities}~\cite{eltabakh2023cross,wang2026unified}, whose distinct structures and representations require modality-specific techniques, while some search needs further span multiple modalities.}

To address the aforementioned challenges, open dataset search has gradually evolved beyond traditional keyword--metadata matching. Two notable trends characterize this evolution: the emergence of increasingly diverse dataset search paradigms for different search needs, and the growing integration of LLMs throughout the dataset search process.

\vspace{2.5pt}
\noindent\textbf{Diversification of Dataset Search Paradigms.}
As users' data needs become more diverse, dataset relevance is no longer limited to whether metadata matches a textual query. Recent studies have developed different search paradigms according to what users seek and how relevance is defined. \emph{Content-aware dataset search} directly uses dataset contents, alone or together with metadata, as search evidence to estimate relevance to textual or exemplar-based inputs~\cite{zhang2018ad,chen2023dense}. \emph{Dataset relationship search} identifies candidate datasets that can establish a useful relationship with a query dataset or data object, such as through joinability, unionability, or correlation~\cite{dong2023deepjoin,khatiwada2023santos,santos2022sketch}. \emph{Task-oriented dataset search} evaluates candidate datasets according to their suitability or utility for a user-specified task~\cite{viswanathan2023datafinder,galhotra2023metam}. These paradigms reflect the expansion of dataset search from simple topical matching toward richer notions of relevance determined by users' actual data needs.

\vspace{2.5pt}{
\noindent\textbf{LLM for Dataset Search.}
Another important trend is the growing application of LLMs to dataset search. LLMs can derive richer searchable evidence from dataset contents, metadata, and external contexts by generating descriptions, extracting semantic relations, and organizing repository-level knowledge~\cite{zhang2026autoddg,wei2026kats,chenarides2026data}. They can also transform dataset information into search-oriented representations, generate retrieval supervision, and support semantic or generative indexing~\cite{silva2024improving,terrenzi2026piper,guo2025birdie,zhang2026generative}. On the user side, LLMs can interpret underspecified data needs, interactively clarify or refine search requirements, decompose complex requests into retrieval-oriented specifications, and plan search actions across different sources and retrieval mechanisms~\cite{long2026sciencedb,wetzel2026query,publicagent2026,yang2025quest2dataagent,li2026octopus}. At the relevance-estimation stage, LLMs can reason over candidate contents, structures, sets, and multiple modalities to validate or rerank retrieved datasets~\cite{soliman2026coret,xu2026efficient,yu2026bringing}. Overall, these applications demonstrate the potential of LLMs to address challenges that are difficult to handle with conventional dataset search techniques.}

These advances have produced a growing body of work across diverse directions of open dataset search. A systematic and comprehensive review is therefore timely to organize recent progress, clarify the landscape, and identify remaining challenges.

\subsection{Survey Scope and Contributions}
\label{sec:survey_scope}
{
Existing surveys have established important foundations for dataset search, with representative studies focusing on Web and tabular data search~\cite{zhang2020web}, metadata-based dataset search~\cite{chapman2020dataset}, broader dataset discovery and exploration~\cite{paton2023dataset}, table discovery in data lakes~\cite{fan2023table}, and more recently LLM-enabled data discovery and integration~\cite{DBLP:journals/debu/FreireFFKLPSSW25}. As open dataset search continues to evolve toward more diverse search paradigms and deeper integration with LLMs, a broader and updated view of the field is needed. Accordingly, this survey provides an updated view of open dataset search spanning its practical foundations, recent search techniques, LLM-enhanced search, and evaluation benchmarks, while focusing the main technical review on advances beyond conventional keyword--metadata matching. We organize recent techniques according to different search needs and notions of relevance and systematically examine how LLMs enhance different stages of the dataset search process. Specifically, our main contributions are as follows:}

\begin{itemize}

    \item {We connect recent research with practical dataset discovery by reviewing the foundations of open dataset search, including sharing and discovery infrastructures, metadata standards, and representative search engines and platforms.}

    \item {We provide a comprehensive review of recent open dataset search techniques across three major search paradigms: content-aware dataset search, dataset relationship search, and task-oriented dataset search.}

    \item {We systematically examine how LLMs enhance dataset search across four stages: search evidence construction, dataset representation and indexing, user input and search planning, and relevance estimation.}

    \item {We review representative dataset search benchmarks and evaluation practices, identify their current limitations, and discuss open issues and promising research directions, including generative, privacy-preserving and federated, quality- and trust-aware, and preparation- and augmentation-oriented dataset search.}

\end{itemize}

{The remainder of this survey is organized as follows. Section~\ref{sec: methodology} describes our survey methodology, including the research questions, literature search and selection process, and data extraction procedure. Section~\ref{sec: systems} reviews the foundations of open dataset search, including dataset sharing and discovery infrastructures and representative search engines and platforms. Section~\ref{sec:ds} reviews recent open dataset search techniques across different search paradigms. Section~\ref{sec:llm_for_dataset_search} examines how LLMs enhance different stages of dataset search. Section~\ref{sec:benchmarks} reviews representative benchmarks and evaluation practices for dataset search. Section~\ref{sec: Future Directions and Open Issues} presents open issues and promising future directions. Section~\ref{sec: Conclusion} concludes this survey.}

\end{sloppypar}

%% file: sections/sec2.tex
\begin{sloppypar}

\section{Survey Methodology}
\label{sec: methodology}

To ensure a systematic and reproducible review process, we follow the principles of systematic literature reviews~\cite{page2021prisma,kitchenham2007guidelines} to identify, select, and analyze relevant studies on open dataset search. This section describes the methodology adopted in this survey, including the research questions, literature search strategy, paper selection criteria, and data extraction process. The identified studies are analyzed according to a unified set of dimensions that characterize dataset search methods, including dataset modalities, search evidence, representations, user inputs, relevance estimation, indexing and acceleration techniques, and evaluation protocols. These dimensions provide the basis for the taxonomy, comparative analysis, and synthesis presented in the following sections.

\subsection{Research Questions}

To guide our review, we formulate the following research questions.

\begin{itemize}
    \item {RQ1: What foundations, infrastructures, and systems support open dataset search?}

    \item {RQ2: What search paradigms and technical approaches have been developed for open dataset search?}

    \item {RQ3: How do large language models enhance different stages of open dataset search?}

    \item {RQ4: What benchmarks and evaluation methodologies have been developed for open dataset search?}

    \item {RQ5: What challenges and future directions remain for open dataset search?}
\end{itemize}
These research questions cover the major topics discussed in this survey, including the foundations of open dataset search, dataset search techniques, LLM-enhanced dataset search, benchmarks and evaluation, and future directions.

\subsection{Literature Search Strategy}

We conducted literature searches using several academic search sources, including ACM Digital Library, IEEE Xplore, SpringerLink, ScienceDirect, VLDB Endowment, Google Scholar, and arXiv. These sources cover major venues in databases, data management, information retrieval, machine learning, and artificial intelligence, where research on dataset search and discovery is commonly published.

The search process focused on studies related to open dataset search and dataset discovery. We constructed search queries by combining terms describing dataset objects, search processes, and relevant tasks or technical approaches. Specifically, the search terms include:

\begin{itemize}
    \item \textbf{Dataset-related terms:}
    ``dataset'', ``set'', ``data'', ``table'', ``data lake'', ``data repository'',
    ``data portal'', ``data catalog'', ``data market'', ``open data'',
    ``structured data'', ``multimodal data''.

    \item \textbf{Search-related terms:}
    ``search'', ``discovery'', ``retrieval'', ``recommendation'',
    ``matching'', ``ranking'', ``exploration'', ``query'', ``information retrieval''.

    \item \textbf{Technique- and task-related terms:}
    ``similarity'', ``join'', ``union'', ``correlation'', ``integration'', ``relationship'', ``completion'', ``semantic'',
    ``embedding'', ``vector'', ``natural language'',
    ``large language model (LLM)'', ``agent''.
\end{itemize}

We combined at least one dataset-related term with one search-related term, and further incorporated technique- or task-related terms when searching for specific research directions. Representative combinations include ``dataset search'' and ``data discovery''. To further improve coverage, we also performed backward and forward citation searches on relevant papers to identify additional studies not retrieved by the initial searches.

\subsection{Paper Selection Criteria}

After collecting candidate studies through the above search strategy, we performed a multi-stage screening process to identify studies relevant to open dataset search. We distinguish between background studies and primary studies in the selection process. Background studies are included to provide necessary context, introduce relevant concepts, or support the discussion of the evolution and ecosystem of open dataset search. Such studies are not required to propose dataset search methods, but their investigated topics should be directly relevant to the corresponding discussions in this survey. Primary studies refer to studies that constitute the main technical body reviewed in this survey; the following criteria were applied during the screening process:

\begin{itemize}
    \item[\checkmark] Studies published between 2022 and 2026, including
    preprints in 2025 and 2026, were considered.

    \item[\checkmark] The research problems investigated or addressed by the
    studies conform to Definition~\ref{def: Open Dataset Search}.

    \item[\checkmark] For studies involving LLMs, the information, representations, or decisions generated by LLMs must be explicitly utilized in at least one stage of the dataset search framework introduced in Section~\ref{subsec: Dataset and Dataset Search}.

    \item[\checkmark] Studies published in high-quality conferences and journals,
such as SIGMOD, VLDB, ICDE, KDD, SIGIR, TODS, and other relevant database, information retrieval, machine learning, and artificial intelligence venues.

    \item[$\times$] Studies that only adopt traditional keyword--metadata matching as the dataset search paradigm were excluded.

    \item[$\times$] Studies focusing only on dataset construction, cleaning,
    transformation, or preparation without discussing dataset search or
    discovery were excluded.

    \item[$\times$] Studies whose actual experimental problem setting is
    retrieving individual data instances or records rather than datasets as
    search targets were excluded, even if the retrieved instances are
    represented as sets and the problem is described as set search.

    \item[$\times$] Duplicate publications or preliminary versions of the
    same work were excluded.
\end{itemize}
After applying the above criteria, we selected 87 primary studies for further
analysis, which constitute the main technical body reviewed in this survey.

\subsection{Data Extraction and Analysis}\label{sec: Data Extraction and Analysis}

Based on the dataset search framework defined in
Section~\ref{subsec: Dataset and Dataset Search}, we extract and analyze
the selected studies along the following dimensions to provide a consistent
basis for comparison. Specifically, \emph{Dataset modality} specifies the
type of datasets being searched, such as tabular, spatial, graph, and
multimodal datasets. \emph{Search evidence} describes the information used
to assess relevance between user inputs and candidate datasets, such as
metadata or dataset contents. \emph{Representation} describes how such
evidence is transformed into a form suitable for relevance estimation and
retrieval, such as semantic embeddings, statistical distributions, sketches,
or graphs. \emph{User input} specifies how users express their dataset needs,
such as keywords, natural-language queries, exemplar datasets, or task
specifications. \emph{Relevance estimation} describes how the relevance
between user inputs and candidate datasets is quantified, such as semantic
similarity, relationship compatibility, dataset-level distance, or downstream
utility. Finally, \emph{Index/acceleration} describes the indexing structures
and optimization techniques used to improve search efficiency, such as
inverted indexes and vector indexes. Based on these extracted dimensions, we perform a thematic analysis to organize existing dataset search techniques according to their search
objectives and relevance definitions. Specifically, we categorize existing
approaches into three major paradigms: content-aware dataset search, dataset
relationship search, and task-oriented dataset search.

For studies involving large language models, we additionally extract
information about the \textit{LLM inputs}, \textit{LLM outputs}, and
\textit{how LLM-generated information is integrated into the dataset search
pipeline}. Based on these dimensions, we organize LLM-enhanced dataset
search methods according to the stages where LLMs are integrated, including
search evidence construction, dataset representation and indexing, user input
and search planning, and relevance estimation. For benchmark studies, we
further extract information about \textit{benchmark scale},
\textit{evaluation target}, \textit{ground-truth construction}, and
\textit{evaluation metrics and protocols}. These extracted characteristics
are used to summarize existing benchmark designs and analyze the current
evaluation landscape of open dataset search.

\section{Foundations of Open Dataset Search}\label{sec: systems}

This section introduces the foundations that support open dataset search in
practice. We first discuss the infrastructures for dataset sharing and
discovery, including repositories, portals, and metadata standards, and then
review representative dataset search engines and platforms that provide
user-facing discovery services.

\vspace{2.5pt}
\noindent\textbf{Dataset Sharing and Discovery Infrastructures.} Open dataset search relies on infrastructures that enable datasets to be shared, organized, and discovered beyond their original producers. These infrastructures determine how datasets are published, described, accessed,
and exposed to search systems. Existing studies have emphasized that dataset
discovery is closely connected with data sharing ecosystems~\cite{castro2024responsible}, including open
data portals, scientific repositories, and data management platforms. Open data portals and
repository systems typically maintain dataset records with descriptive
metadata and provide basic discovery functions, such as keyword search,
filtering, and browsing~\cite{chapman2020dataset,paton2023dataset}.

Beyond publicly accessible repositories, dataset sharing has expanded toward
more diverse ecosystems, including data markets and controlled-access data
platforms. These infrastructures support dataset exchange under additional
requirements such as ownership, licensing, access policies, and governance
mechanisms~\cite{zhang2024survey}. Meanwhile, responsible data sharing
principles, such as the FAIR principles~\cite{wilkinson2016fair}, emphasize
that datasets should be findable, accessible, interoperable, and reusable
through mechanisms such as persistent identifiers and rich metadata.
Data catalogs organize dataset records and expose their metadata for search,
while metadata standards improve interoperability across heterogeneous
repositories. For example, schema.org~\cite{brickley2019google} and
DCAT~\cite{albertoni2024w3c} provide structured descriptions for dataset
discovery across Web pages and catalogs, while
Croissant~\cite{akhtar2024croissant} introduces a metadata format
specifically designed for machine learning datasets to improve dataset
portability and discoverability.

\vspace{2.5pt}
\noindent\textbf{Dataset Search Engines and Platforms.} While sharing infrastructures and metadata standards provide the foundation for dataset discovery, dataset search engines and platforms operationalize these capabilities by indexing datasets and providing user-facing search
interfaces. Table~\ref{tab: dataset search systems} summarizes representative dataset search engines. For each system, we report its supported query mechanisms, user upload capability,
open-source availability, and API access.

\begin{table}[h]
	\centering
	\caption{Examples of dataset search engines across domains.}
	\label{tab: dataset search systems}
	\scalebox{0.8}{
	\begin{tabular}{
    >{\raggedright\arraybackslash}p{2.5cm} >{\raggedright\arraybackslash}p{3.2cm} >{\raggedright\arraybackslash}p{3.2cm} >{\centering\arraybackslash}p{1.4cm} >{\centering\arraybackslash}p{1.2cm} >{\centering\arraybackslash}p{1.2cm}}
		\toprule
        \addlinespace[-0.2pt]
		\rowcolor{gray!30}
		\textbf{Dataset Domain} & \textbf{Search Engine} & \textbf{Query Mechanism} & \textbf{User Upload}  & \textbf{Open Source} & \textbf{API} \\
		[-2pt]\midrule

		 &
		Google Dataset Search~\cite{brickley2019google} &
		Keywords &
		No &
		No &
		Limited \\
		
		Generic &
		CKAN-based Portal$^1$ &
		Keywords, Faceted &
		Yes &
		Yes &
		Yes \\
		
	    &
		Data.world$^2$ &
		Keywords, SQL &
		Yes &
		Partial &
		Yes \\

        \addlinespace[0pt]\midrule

		&
		IEEEDataPort$^3$ &
		Keywords &
		Yes &
		No &
		Yes \\

		&
		PapersWithCode$^4$ &
		Keywords, Task-based &
		Yes &
		Partial &
		Yes \\

		&
		Kaggle$^5$ &
		Keywords, Tags &
		Yes &
		No &
		Limited \\

		Academic &
		Huggingface Datasets$^6$ &
		Keywords, Task-based &
		Yes &
		Yes &
		Yes \\

		&
		Delve~\cite{akujuobi2017delve} &
		Keywords &
		No &
		No &
		No \\

		&
		DataLab~\cite{xiao2022datalab} &
		Task-based &
		No &
		Yes &
		Yes \\

         \addlinespace[0pt]\midrule
		
		&
		GeoBlacklight$^7$ &
		Keywords, Spatial filters &
		No &
		Yes &
		Yes \\
		
		 &
		UCR STAR~\cite{ghosh2019ucr} &
		Keywords &
		Yes &
		No &
		No \\

        Geospatial&
        Data.gov GeoPlatform$^8$ &
        Keywords, Faceted, Spatial filters &
        No &
        No &
        Yes \\

        &
        OpenStreetMap$^9$ &
        Region-based, Tags &
        No &
        No &
        No \\
		
        \addlinespace[0pt]\midrule

         &
        NCBI Datasets$^{10}$ &
        Keywords, ID-based &
        No &
        No &
        Yes \\

        &
        EBI BioStudies$^{11}$ &
        Keywords &
        Yes &
        No &
        Yes \\

        Biomedical&
        PhysioNet$^{12}$ &
        Keywords, Tags &
        Partial &
        Yes &
        Yes \\

        &
        TCIA$^{13}$ &
        Keywords &
        No &
        No &
        Yes \\

		\bottomrule

		\multicolumn{6}{l}{$^1$\url{https://ckan.org/}, 
		$^2$\url{https://data.world/}, 
		$^3$\url{https://ieee-dataport.org/}, 
		$^4$\url{https://paperswithcode.com/},}\\
		\multicolumn{6}{l}{
		$^5$\url{https://www.kaggle.com/datasets}, 
		$^6$\url{https://huggingface.co/datasets},
        $^7$\url{https://geoblacklight.org/about/},
        } \\
        \multicolumn{6}{l}{
		$^8$\url{https://data.geoplatform.gov/}, 
		$^9$\url{https://download.geofabrik.de/},
        $^{10}$\url{https://www.ncbi.nlm.nih.gov/datasets/},
        } \\
        \multicolumn{6}{l}{
        $^{11}$\url{https://www.ebi.ac.uk/biostudies/},
        $^{12}$\url{https://physionet.org/},
        $^{13}$\url{https://www.cancerimagingarchive.net/}.
        } \\
	\end{tabular}

	}
\end{table}

\renewcommand{\thefootnote}{}
\footnotetext{Unless otherwise specified, ``dataset search''  refers to ``open dataset search'' in the following text.}%
\addtocounter{footnote}{-1}
\renewcommand{\thefootnote}{\arabic{footnote}}

Generic dataset search engines aim to provide broad coverage across diverse
dataset collections and domains. These systems typically rely on dataset
metadata and support modality-agnostic discovery. For example, Google Dataset
Search~\cite{brickley2019google} indexes datasets exposed through Web pages
using schema.org markup and supports keyword-based queries over fields such as
titles, descriptions, and publishers. CKAN-based portals and
Data.world further provide metadata-based search with filtering and
faceted navigation, allowing users to discover datasets according to
attributes such as format, organization, or topic.

Academic-oriented dataset search platforms focus on supporting research and
machine learning workflows by connecting datasets with publications,
benchmarks, or computational tasks. Systems such as IEEEDataPort and
PapersWithCode organize datasets associated with scientific publications
and machine learning tasks, respectively. Kaggle and HuggingFace
Datasets additionally provide dataset hosting, user contributions, and
programmatic access through APIs. Beyond general dataset repositories,
platforms such as DataLab~\cite{xiao2022datalab} support task-oriented
exploration of language datasets, while Delve~\cite{akujuobi2017delve}
facilitates dataset discovery through scholarly information and citation
analysis.

Domain-specific dataset search engines further adapt discovery mechanisms to
the characteristics of particular data types and application scenarios.
Geospatial platforms, such as GeoBlacklight and Data.gov GeoPlatform,
support spatial filters and map-based exploration for discovering
location-related datasets, while OpenStreetMap Extracts enables
region-based access to geographic data. Biomedical platforms address the
specialized requirements of scientific and clinical data discovery. For
example, NCBI Datasets and EBI BioStudies provide access to
biological datasets through keyword and identifier-based search, whereas
PhysioNet and The Cancer Imaging Archive (TCIA) support the
discovery of physiological and medical imaging datasets through domain-specific
metadata and access mechanisms.

\vspace{2.5pt}
\noindent\textbf{Discussion.}~
Existing dataset sharing infrastructures and search platforms provide the foundation for open dataset search by making datasets available, describable, and accessible through metadata-driven discovery mechanisms. However, most existing systems still rely heavily on keyword and metadata
matching, which is scalable and widely adopted but may be insufficient for complex dataset needs. Metadata alone often provides only coarse descriptions and fails to capture dataset contents, relationships, or usefulness for downstream tasks. To address these limitations, recent
research has explored richer search evidence and more diverse notions of dataset relevance, including content-aware, relationship-based, and task-oriented dataset search paradigms, which are discussed in the following sections.

\end{sloppypar}

%% file: sections/sec3.tex
\begin{sloppypar}

\section{Dataset Search Techniques}\label{sec:ds}

This section reviews existing dataset search techniques through three major search paradigms, distinguished primarily by their search objectives and
notions of relevance. Following the analysis framework introduced in
Section~\ref{sec: Data Extraction and Analysis}, we compare representative
methods along six dimensions: \emph{dataset modality}, \emph{search evidence},
\emph{representation}, \emph{user input}, \emph{relevance estimation}, and
\emph{index/acceleration}.

\subsection{Content-aware Dataset Search}

Content-aware dataset search arises from a fundamental limitation of
metadata-based dataset search: metadata alone often provides only a coarse
description of a dataset and may fail to expose the entities, attributes,
values, facts, or other information contained within it. Existing studies
therefore exploit dataset contents as additional search
evidence~\cite{zhang2018ad}. The defining characteristic of this paradigm is
that dataset contents, rather than metadata alone, directly participate in
representing datasets and estimating their relevance to users' search needs.
Early studies mainly focused on tabular data, exploiting structured contents
such as schemas, values, entities, rows, columns, and cells as search evidence,
and developing semantic, vector-based, or structure-aware representations to
support retrieval~\cite{zhang2018ad,deng2019table2vec,chen2020table,
shraga2020web,shraga2020ad,wang2021retrieving}. Recent studies further extend
content-aware search to other forms of dataset contents, such as spatial
points and their distributions. Across these settings, two central questions
arise: \emph{what dataset contents should be used as search evidence}, and
\emph{how such evidence should be represented and matched for relevance
estimation}.

Recent studies mainly develop content-aware dataset search along two
complementary directions. Some select, organize, or summarize dataset
contents into search-oriented evidence, such as snippets, chunks, summaries,
or structured fields, while others focus on how different forms of content
can be represented and matched against textual or exemplar-based inputs.
Accordingly, we review existing studies from these two perspectives; related
papers are summarized in Table~\ref{tab:content_aware_dataset_search}.

\begin{table*}[t]
\centering
\caption{Summary of representative methods for content-aware dataset search.}
\label{tab:content_aware_dataset_search}
\tiny
\setlength{\tabcolsep}{2pt}
\renewcommand{\arraystretch}{0.95}

\resizebox{0.995\textwidth}{!}{%
\begin{tabular}{
>{\raggedright\arraybackslash}p{1.45cm}
>{\raggedright\arraybackslash}p{1.00cm}
>{\raggedright\arraybackslash}p{2.25cm}
>{\raggedright\arraybackslash}p{1.90cm}
>{\raggedright\arraybackslash}p{1.25cm}
>{\raggedright\arraybackslash}p{2.10cm}
>{\raggedright\arraybackslash}p{1.90cm}
}
\hline
\rowcolor{gray!30}
\textbf{Method} &
\textbf{Modality} &
\textbf{Search Evidence} &
\textbf{Representation} &
\textbf{User Input} &
\textbf{Relevance Estimation} &
\textbf{Index / Acceleration} \\
\hline

\rowcolor{gray!12}
\multicolumn{7}{l}{\textit{Content Evidence Construction}} \\
\hline
\papersep

CDS~\cite{chen2024enhancing}
& TAB / RDF / DOC
& Metadata / data snippets
& Snippet vectors
& Keywords
& Learned matching
& Vector index + Pruning \\

DUNKS~\cite{chen2024dunks}
& TAB / RDF / DOC
& Metadata / chunk summaries
& Chunk-summary vectors
& Keywords
& Learned matching
& Chunking + Summarization \\

$\mu$DS~\cite{zhou2025muds}
& TAB / RDF / DOC
& Metadata / data snippets
& Snippet vectors
& Keywords
& Learned matching
& Approximation \\

Caddie~\cite{wang2026caddie}
& RDF
& Metadata / RDF fields
& Fielded sparse vectors
& Keywords
& Weighted BM25
& Inverted index \\
\papersep

\hline
\rowcolor{gray!12}
\multicolumn{7}{l}{\textit{Content Representation and Matching}} \\
\hline
\papersep

Chen et al.~\cite{chen2023dense}
& RDF
& Metadata / RDF triples
& Dataset vectors
& Keywords
& Dot product
& Inverted index \\
\papersep

Solo~\cite{wang2023solo}
& TAB
& Table title / column names / cell values
& Table / triple vectors
& NL
& Learned matching + max pooling
& Vector index \\

ConTR~\cite{wu2025retrieving}
& TAB
& Headers / row values
& Table-segment vectors
& NL
& Cosine similarity + max pooling
& Vector index \\

THYME~\cite{li2025tailoring}
& TAB
& Titles / headers / cell values
& Field vectors
& NL
& Dense / sparse field matching
& Vector + inverted indexes \\
\papersep

TabRAG~\cite{xu2026efficient}
& TAB
& Headers / cell values
& Rendered table images
& NL
& Cross-modal matching + reranking
& Vector index \\

Spadas~\cite{yang2024unified}
& SPA
& Spatial points
& Point sets
& Spatial exemplar
& Hausdorff distance / overlap
& Tree index + Pruning \\
\papersep

DBF~\cite{yang2022fast}
& SPA
& Spatial points
& Z-order histograms
& Spatial exemplar
& EMD
& Tree index + Bounds \\
\papersep

PriDAS~\cite{li2024privacy}
& SPA
& Spatial points
& Density grids
& Spatial exemplar
& Density overlap
& Two-level index + Pruning \\
\papersep

GIDS~\cite{zhang2026grayscale}
& SPA
& Spatial points
& Grayscale images
& Spatial exemplar
& Image similarity
& Tree index + Bounds \\
\papersep

Fainder~\cite{behme2024fainder}
& TAB
& Attribute values
& Distribution summaries
& Tabular exemplar
& Distribution similarity
& Distribution-based index \\

\hline
\end{tabular}%
}

\vspace{1pt}
\begin{minipage}{0.98\textwidth}
\tiny
\textit{Note.}
\textbf{Modality:}
TAB = Tabular;
RDF = Resource Description Framework;
DOC = Document;
SPA = Spatial.
\textbf{User Input:}
NL = Natural Language.
\textbf{Relevance Estimation:}
BM25 = Best Matching 25;
EMD = Earth Mover's Distance.
\end{minipage}
\end{table*}

\vspace{2.5pt}
\noindent\textbf{Content Evidence Construction.}
Directly using complete dataset contents as search evidence is often
impractical because datasets can be large, heterogeneous, and highly
structured. A line of work therefore selects, structures, or summarizes
dataset contents into search-oriented evidence.
CDS~\cite{chen2024enhancing} extracts query-biased snippets from
heterogeneous structured datasets. DUNKS~\cite{chen2024dunks} transforms
large contents into searchable chunks and summaries, while
$\mu$DS~\cite{zhou2025muds} formulates snippet construction as a
multi-objective problem. Caddie~\cite{wang2026caddie} incorporates metadata
and RDF fields into fielded search evidence for end-to-end dataset retrieval.
Together, these methods reduce reliance on complete raw contents by
constructing selected, summarized, or structured evidence for search.

\vspace{2.5pt}
\noindent\textbf{Content Representation and Matching.}
Given dataset contents as search evidence, another line of work investigates
how to represent such evidence and match it against user inputs. For RDF
datasets, Chen et al.~\cite{chen2023dense} select representative triples,
transform them into textual representations, and apply dense matching to
rerank candidates. For tabular datasets, recent studies adopt fine-grained
representations and matching mechanisms tailored to different table
components. Solo~\cite{wang2023solo} exploits titles, columns, and cell values
through semantic representations and learned matching, while
ConTR~\cite{wu2025retrieving} and THYME~\cite{li2025tailoring} further model
row-level and field-level semantics using contrastive learning and hybrid
matching. TabRAG~\cite{xu2026efficient} extends table representation to visual
features and performs cross-modal matching with reranking.

Content-aware search can also use an exemplar dataset as the user input.
For spatial datasets, Spadas~\cite{yang2024unified} represents datasets as
point sets and compares them using measures such as Hausdorff distance and
spatial overlap. DBF~\cite{yang2022fast} summarizes spatial distributions as
Z-order histograms and measures their similarity using Earth Mover's
Distance (EMD). PriDAS~\cite{li2024privacy} represents datasets with
grid-based density distributions and estimates relevance through density
overlap, while GIDS~\cite{zhang2026grayscale} converts spatial distributions
into grayscale images and compares the resulting image representations.
Beyond spatial data, Fainder~\cite{behme2024fainder} captures attribute
distributions and builds distribution-aware indexes to retrieve datasets
with similar statistical characteristics. Together, these studies show that
content-aware dataset search couples dataset representations with relevance
functions adapted to different types of content and user inputs.

\vspace{2.5pt}
\noindent\textbf{Discussion.}
Content-aware dataset search has evolved beyond metadata-only matching by
directly exploiting increasingly fine-grained and heterogeneous dataset
contents, including fields, triples, snippets, table segments, spatial
points, and distributions. However, existing methods still face a trade-off between information preservation and search efficiency.
Using richer or more fine-grained dataset contents can preserve more
information for relevance estimation, but also increases indexing, storage,
and matching costs; conversely, compact snippets, summaries, or coarse
representations improve efficiency but may omit useful evidence or structural
characteristics. More adaptive evidence selection and structure-preserving
representations could better balance these two aspects. In addition, most
existing methods are designed for particular user-input forms and relevance
functions, such as semantic matching for textual queries or distributional
similarity for exemplar datasets. Developing more flexible representations
and relevance estimation mechanisms that support multiple input forms would
improve the generality of content-aware dataset search.

\subsection{Dataset Relationship Search}

\begin{figure*}[t]
\includegraphics[width=0.85\textwidth]{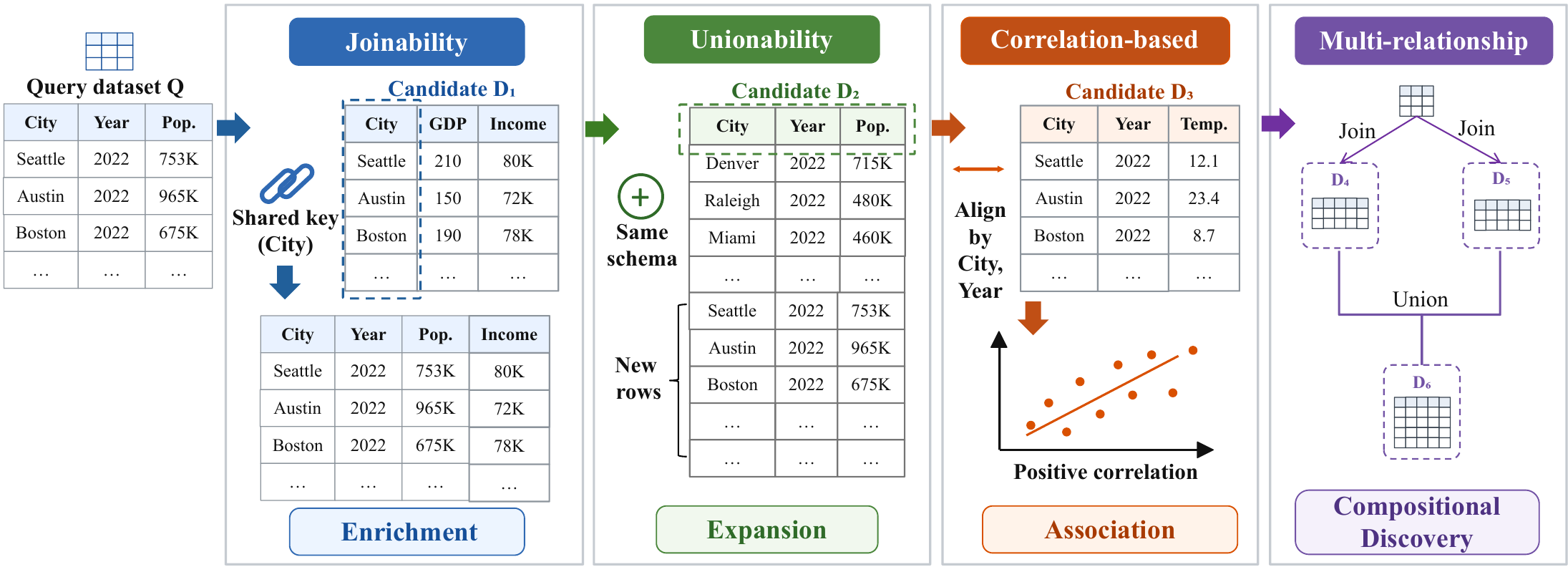}
\caption{Major relationship types between a query dataset and candidate datasets.}
\label{fig: relationship_types}
\end{figure*}

Metadata- or content-aware search can retrieve datasets based on their
descriptions or contents, but many data analysis tasks require a different
notion of relevance: whether a candidate dataset can establish a useful
relationship with an existing data object. Given an existing dataset, users
may seek another dataset that can enrich it through a join, extend it with
compatible records, or reveal meaningful associations after alignment. These
needs motivate dataset relationship search, where a candidate is relevant not
simply because it contains similar information, but because it can establish
a useful relationship with the query dataset. Such relationships include
joinability, unionability, correlation, and more complex relationships
involving multiple datasets or discovery operations. As shown in
Figure~\ref{fig: relationship_types}, we categorize dataset relationships
into four major types:

\begin{itemize}
\item \textbf{Joinability} links datasets through shared or alignable units
to support data enrichment.

\item \textbf{Unionability} identifies compatible datasets that can provide
additional records for data expansion.

\item \textbf{Correlation-based relationship} identifies datasets that
exhibit meaningful statistical associations after alignment.

\item \textbf{Multi-relationship discovery} considers multiple relationship
types or datasets within a unified or compositional discovery process.
\end{itemize}

\begin{table*}[t]
\centering
\caption{Summary of representative methods for dataset relationship search.}
\label{tab:dataset_relationship_search}

\tiny
\setlength{\tabcolsep}{2pt}
\renewcommand{\arraystretch}{0.95}

\resizebox{0.995\textwidth}{!}{%
\begin{tabular}{
>{\raggedright\arraybackslash}p{1.50cm}
>{\raggedright\arraybackslash}p{0.72cm}
>{\raggedright\arraybackslash}p{0.40cm}
>{\raggedright\arraybackslash}p{2.05cm}
>{\raggedright\arraybackslash}p{1.95cm}
>{\raggedright\arraybackslash}p{1.00cm}
>{\raggedright\arraybackslash}p{2.10cm}
>{\raggedright\arraybackslash}p{1.85cm}
}

\hline
\rowcolor{gray!30}
\textbf{Method} &
\textbf{Modality} &
\textbf{Type} &
\textbf{Search Evidence} &
\textbf{Representation} &
\textbf{User Input} &
\textbf{Relevance Estimation} &
\textbf{Index / Acceleration} \\
\hline

\rowcolor{gray!12}
\multicolumn{8}{l}{\textit{Semantic Relationship Representation}} \\
\hline
\papersep

DeepJoin~\cite{dong2023deepjoin}
& TAB
& J
& Column values / names / table title
& Column vectors
& Column
& Euclidean distance
& Vector index \\

WarpGate~\cite{cong2023warpgate}
& TAB
& J
& Column values
& Column vectors
& Column
& Cosine similarity
& LSH + Sampling \\
\papersep

Snoopy~\cite{guo2025snoopy}
& TAB
& J
& Column values
& Column vectors
& Column
& Cosine similarity
& Vector index \\
\papersep

Starmie~\cite{fan2023semantics}
& TAB
& U
& Table values
& Column vectors
& Table
& Cosine + bipartite matching
& Vector index + Pruning \\
\papersep

Pylon~\cite{cong2023pylon}
& TAB
& U
& Column values
& Column vectors
& Table
& Cosine similarity
& LSH \\
\papersep

SANTOS~\cite{khatiwada2023santos}
& TAB
& U
& Column values / same-row values
& Table semantic graphs
& Table + NL
& Semantic matching
& Inverted index \\

AUTOTUS~\cite{hu2023automatic}
& TAB
& U
& Table values / metadata
& Column-pair vectors
& Table
& Learned similarity
& N/A \\
\papersep

LIFTus~\cite{qiu2025liftus}
& TAB
& U
& Column values / statistics
& Column vectors
& Table
& Learned similarity
& Vector index \\
\papersep

EasyTUS~\cite{otto2025easytus}
& TAB
& U
& Table values
& Table vectors
& Table
& Cosine similarity
& Vector index \\
\papersep

GenTUS~\cite{zhang2026generative}
& TAB
& U
& Table values
& Discrete table IDs
& Table
& Generative relevance
& Quantization \\
\papersep

\hline

\rowcolor{gray!12}
\multicolumn{8}{l}{\textit{Relationship-aware Alignment and Scoring}} \\
\hline
\papersep

FREYJA~\cite{maynou2026freyja}
& TAB
& J
& Column statistics
& Statistical profiles
& Column
& Learned similarity
& Profiling \\
\papersep

MosaicJoin~\cite{fan2026mosaicjoin}
& TAB
& J
& Column values
& Column sketches
& Column
& Value-level semantic matching
& Sketching + Sampling \\

PolyJoin~\cite{hu2025polyjoin}
& TAB
& J
& Column values / metadata
& Multi-column vectors
& Table
& Semantic overlap
& N/A \\
\papersep

SemDisc~\cite{mohammad2026qualitative}
& TAB
& J
& Column values
& Join graphs
& Table
& Tuple matching
& Inverted index \\
\papersep

TACTUS~\cite{sun2026efficient}
& TAB
& U
& Table values
& Table + column vectors
& Table
& Cosine + column matching
& Vector index + Pruning \\
\papersep

MSDS~\cite{yang2025joinable}
& SPA
& J
& Spatial coordinates
& Spatial grids
& Spatial dataset
& Spatial overlap / coverage
& Tree index \\

QCR index~\cite{santos2022sketch}
& TAB
& C
& Matching-column / numeric values
& Column-pair sketches
& Table
& Jaccard + correlation
& Sketching + Inverted index \\

TUPSK~\cite{santos2024mutual}
& TAB
& C
& Matching-column / target-column values
& Column-pair sketches
& Table
& Mutual information
& Sketching \\

Nexus~\cite{gong2024nexus}
& SPA
& C
& Spatial / temporal / numeric values
& Spatio-temporal aggregates
& Dataset
& Pearson correlation
& Hash + Inverted index \\

\hline

\rowcolor{gray!12}
\multicolumn{8}{l}{\textit{Multi-signal and Compositional Relationship Discovery}} \\
\hline
\papersep

OmniMatch~\cite{koutras2025omnimatch}
& TAB
& J
& Column values
& Multi-relational column graph
& Dataset collection
& Learned column matching
& N/A \\

TabSketch\allowbreak{}FM~\cite{khatiwada2025tabsketchfm}
& TAB
& M
& Table values / metadata
& Column vectors
& Table
& Nearest-neighbor similarity
& Vector search \\
\papersep

CMDL~\cite{eltabakh2023cross}
& TAB / DOC
& M
& Document text / column values / metadata
& Joint document--column vectors + sketches
& Document / Column
& Cross-modal similarity + relation composition
& Vector + LSH + inverted indexes \\

Gen-T~\cite{fan2024gent}
& TAB
& M
& Table values
& Value-overlap matrices
& Table
& Error-aware tuple matching
& Pruning \\
\papersep

BLEND~\cite{esmailoghli2025blend}
& TAB
& M
& Table values / same-row values
& Cell-level records
& NL
& Overlap / correlation
& Composite index + Query optimization \\

\hline
\end{tabular}%
}

\vspace{1pt}

\begin{minipage}{0.98\textwidth}
\tiny
\textit{Note.}
\textbf{Modality:}
TAB = Tabular;
DOC = Document;
SPA = Spatial.
\textbf{Type:}
J = Joinability;
U = Unionability;
C = Correlation-based;
M = Multi-relationship.
\textbf{User Input:}
NL = Natural Language.
\textbf{Index / Acceleration:}
LSH = Locality-Sensitive Hashing.
\end{minipage}

\end{table*}

Early studies mainly characterized these relationships through explicit
structural and content compatibility. Joinability was commonly modeled
through value overlap or containment between columns~\cite{zhu2016lsh,
fernandez2019lazo,zhu2019josie}, with later semantic approaches relaxing
exact value matching~\cite{dong2021efficient}. Unionability was traditionally
determined through common subjects, compatible attribute domains, and schema
alignment~\cite{sarma2012finding,nargesian2018table}, while
correlation-based methods evaluated statistical associations after datasets
were aligned over shared keys or spatio-temporal units~\cite{chirigati2016data,
santos2021correlation}. Broader discovery systems further combined schema,
content, and semantic signals to identify related data assets and join
paths~\cite{fernandez2018aurum,bogatu2020dataset}. Together, these studies
established a central technical problem: how to represent and align datasets
so that relationship-specific relevance can be estimated effectively. Recent
studies mainly advance this problem through semantic relationship
representation, relationship-aware alignment and scoring, and multi-signal or
compositional relationship discovery. Related papers are summarized in
Table~\ref{tab:dataset_relationship_search}.

\vspace{2.5pt}
\noindent\textbf{Semantic Relationship Representation.}
A major development in relationship search is the transition from explicit
overlap and schema compatibility toward learned representations that capture
semantic relationships between datasets. This shift is particularly evident
in joinability and unionability search, where syntactically different
contents may nevertheless participate in the same data operation.

For joinability, DeepJoin~\cite{dong2023deepjoin} encodes columns with
pre-trained language models and retrieves joinable columns through
approximate nearest-neighbor search, allowing semantic relationships to be
captured beyond exact value overlap. WarpGate~\cite{cong2023warpgate}
adopts a similar embedding-based formulation for semantic join discovery in
cloud data warehouses, emphasizing interactive discovery across large
enterprise repositories. More recent work further distinguishes generic
semantic similarity from relationship-specific compatibility.
Snoopy~\cite{guo2025snoopy} observes that semantically similar column
embeddings do not necessarily imply strong cell-level joinability and
therefore constructs proxy-column-based representations that approximate
column matching behavior.

Unionability search exhibits a parallel evolution from explicit domain
compatibility toward contextualized and relational representations.
Starmie~\cite{fan2023semantics} learns contextualized column embeddings
through self-supervised contrastive learning, while
Pylon~\cite{cong2023pylon} jointly considers representation learning and
retrieval structure. SANTOS~\cite{khatiwada2023santos} represents tables as
semantic graphs that preserve both column semantics and inter-column
relationships, while AUTOTUS~\cite{hu2023automatic} further models
contextualized column-pair relations. LIFTus~\cite{qiu2025liftus}
complements these methods by learning adaptive multi-aspect column
representations for heterogeneous column contents.

More recently, representations have moved beyond conventional column
embeddings. EasyTUS~\cite{otto2025easytus} serializes entire tables into
semantic vectors for table-level retrieval, whereas
GenTUS~\cite{zhang2026generative} represents tables with unionability-aware
semantic identifiers and directly generates ranked candidates through
constrained decoding. Overall, relationship representation has evolved from
generic embeddings toward representations that better capture relational
structure, heterogeneous contents, and relationship-specific semantics.

\vspace{2.5pt}
\noindent\textbf{Relationship-aware Alignment and Scoring.}
Representing dataset semantics alone does not determine whether two datasets
can participate in a desired relationship. Relationship search must also
identify the units through which datasets should be aligned and evaluate
compatibility under the corresponding operation. Recent studies therefore
design alignment and scoring mechanisms tailored to specific relationship
structures.

For joinability, FREYJA~\cite{maynou2026freyja} combines lightweight column
profiles with relationship-specific scoring to estimate join quality.
MosaicJoin~\cite{fan2026mosaicjoin} instead uses compact semantic sketches to
preserve fine-grained value-level matching while avoiding exhaustive
comparisons between high-cardinality columns. PolyJoin~\cite{hu2025polyjoin}
extends the alignment unit from individual columns to multi-column keys,
while SemDisc~\cite{mohammad2026qualitative} further moves beyond direct
column pairs by searching join paths whose materialized results satisfy
query examples. These methods progressively expand joinability estimation
from column-level compatibility toward value-level, composite-key, and
path-level alignment.

Unionability also depends on the granularity at which compatibility is
assessed. TACTUS~\cite{sun2026efficient} challenges the dominant
column-centric pipeline, observing that a few compatible columns do not
necessarily imply that two tables are globally unionable. It therefore first
evaluates candidates using holistic table representations and subsequently
combines table-level and column-level evidence for reranking. This
table-first formulation illustrates that the choice of alignment granularity
directly affects the resulting notion of relationship relevance.

Relationship-specific alignment becomes even more explicit when the shared
units are not conventional table keys. MSDS~\cite{yang2025joinable}
formulates joinability over multi-source spatial datasets, where
relationships depend on spatial intersection, coverage gain, and
connectivity rather than column-value correspondence. Correlation-based
search similarly requires datasets to be aligned before their relevance can
be assessed. Santos et al.~\cite{santos2022sketch} jointly encode join-key
overlap and aligned numeric variation into retrieval-oriented sketches,
allowing candidates to be retrieved according to both joinability and
potential correlation. TUPSK~\cite{santos2024mutual} further uses tuple
sketches to estimate mutual information between attributes across joinable
tables without materializing their joins, enabling candidates to be ranked
by stronger post-join statistical dependence. Nexus~\cite{gong2024nexus}
extends relationship discovery to collections of spatio-temporal tables by
aligning datasets over compatible spatial and temporal units and computing
correlations over the resulting aggregates. Together, these studies show
that alignment and statistical association are becoming integral to
relationship-aware candidate retrieval rather than being treated solely as
post-retrieval analysis.

\vspace{2.5pt}
\noindent\textbf{Multi-signal and Compositional Relationship Discovery.}
Most relationship-specific methods optimize one predefined notion of
relevance. Practical discovery tasks, however, may depend on multiple
relationship signals, involve several relationship types, or require
multiple datasets to jointly satisfy a search need. Recent work therefore
explores representations and search frameworks that move beyond isolated
pairwise relationships.

One direction combines heterogeneous signals to infer a target relationship.
OmniMatch~\cite{koutras2025omnimatch} constructs a multi-relational column
graph from value overlap, token similarity, embedding similarity, and
distributional similarity, and uses graph-based learning to infer joinable
column pairs. Rather than relying on one predetermined similarity function,
it exploits interactions among complementary relationship signals. Another
direction seeks representations that can be reused across several
relationship types. TabSketchFM~\cite{khatiwada2025tabsketchfm} learns
sketch-enhanced table representations and adapts them to joinability,
unionability, and subset or superset search, demonstrating that some
relationship evidence can be shared across discovery tasks rather than
encoded through entirely separate indexes and models.
CMDL~\cite{eltabakh2023cross} further extends relationship discovery across
structured and unstructured data lakes. It learns shared representations of
documents and table columns to identify cross-modal semantic relationships
and composes these relationships with structured relationships such as
joinability and unionability.

Relationship discovery can also become compositional at the result level.
Gen-T~\cite{fan2024gent} formulates table reclamation as searching for a set
of originating tables whose contents can collectively reconstruct a query
table. Here, the relevance of a candidate cannot be fully determined
independently, since its contribution depends on the other tables selected
for the reconstruction objective. BLEND~\cite{esmailoghli2025blend}
provides a complementary operator-oriented formulation, supporting join,
multi-column join, union, correlation, and other discovery operations
through a unified relational index and allowing these operators to be
composed into more complex discovery plans. These approaches indicate a
broader transition from independently searching for one predefined pairwise
relationship toward jointly modeling, reusing, and composing multiple forms
of inter-dataset relevance.

\vspace{2.5pt}
\noindent\textbf{Discussion.}
Dataset relationship search has progressively moved from explicit value and
schema compatibility toward semantic, alignment-aware, and compositional
notions of relevance. Several challenges follow directly from this
relationship-based notion of relevance. First, relationship evidence is
inherently \emph{operation-dependent}. A representation that preserves
joinability does not necessarily preserve unionability or correlation, and
some relationships are directional, making it difficult to construct
reusable dataset representations without obscuring relationship-specific
signals. Second, \emph{alignment is part of relevance estimation itself}.
Alternative key correspondences, schema mappings, spatial or temporal
granularities, and join paths can produce fundamentally different
relationships between the same datasets, so relevance cannot always be
determined before the appropriate alignment is identified. Finally,
relationship relevance is increasingly \emph{compositional rather than
pairwise}. Join paths, table reclamation, and multi-operator discovery depend
on interactions among several datasets, whereas conventional retrieval
typically scores and ranks each candidate independently. Capturing such
alignment-dependent and collective relevance remains a central challenge of
this paradigm.

\subsection{Task-oriented Dataset Search}

Dataset search is increasingly driven not only by what data contain, but also
by what users intend to accomplish with the retrieved
datasets~\cite{viswanathan2023datafinder,galhotra2023metam}. Such scenarios
motivate \emph{task-oriented dataset search}, which evaluates candidate
datasets according to their suitability or utility for a user-specified task.
Unlike content-aware or relationship search, task-oriented search defines
relevance with respect to the intended use of the retrieved data, regardless
of the form of user input. For example, ``Find datasets that improve my sales
forecasting model'' requires data that benefit the specified prediction task.
Content relevance or inter-dataset relationships may help identify promising
candidates, but do not by themselves establish such task-specific benefits.

Existing studies mainly instantiate task-oriented dataset search in two
forms. \emph{Task-aware dataset recommendation} retrieves datasets according
to their suitability for a research or analytical need, whereas
\emph{downstream utility-driven dataset search} evaluates or estimates how
candidate data affect an executable downstream task, such as whether they
improve model performance~\cite{huang2024mileena,zhu2025toast}. These two
forms reflect different levels of task-specific relevance estimation.
Table~\ref{tab:task_oriented_dataset_search} summarizes representative
methods.

\begin{table*}[t]
\centering
\caption{Summary of representative methods for task-oriented dataset search.}
\label{tab:task_oriented_dataset_search}

\tiny
\setlength{\tabcolsep}{2pt}
\renewcommand{\arraystretch}{0.95}

\resizebox{0.995\textwidth}{!}{%
\begin{tabular}{
>{\raggedright\arraybackslash}p{1.25cm}
>{\raggedright\arraybackslash}p{0.75cm}
>{\raggedright\arraybackslash}p{2.65cm}
>{\raggedright\arraybackslash}p{2.10cm}
>{\raggedright\arraybackslash}p{1.45cm}
>{\raggedright\arraybackslash}p{2.20cm}
>{\raggedright\arraybackslash}p{2.00cm}
}

\hline
\rowcolor{gray!30}
\textbf{Method} &
\textbf{Modality} &
\textbf{Search Evidence} &
\textbf{Representation} &
\textbf{User Input} &
\textbf{Relevance Estimation} &
\textbf{Index / Acceleration} \\
\hline

\rowcolor{gray!12}
\multicolumn{7}{l}{\textit{Task-aware Dataset Recommendation}} \\
\hline
\papersep

DataFinder~\cite{viswanathan2023datafinder}
& MUL
& Dataset descriptions / metadata / paper titles
& Dataset vectors
& NL / keywords
& Learned semantic matching
& Vector index \\

KATS~\cite{wei2026kats}
& MUL
& Task descriptions / metadata / task--dataset relations
& Task / dataset vectors + KG
& NL
& Cosine + graph propagation + reranking
& Vector index + Graph propagation \\

ScienceDB AI~\cite{long2026sciencedb}
& MUL
& Dataset descriptions / metadata / scientific requirements
& Dataset vectors
& Multi-turn NL
& Dense matching + reranking
& Vector index + Two-stage retrieval \\
\papersep

Yu et al.~\cite{yu2026bringing}
& MUL
& Dataset descriptions / metadata / task--dataset relations
& Dataset vectors + lexical text
& NL
& Dense / lexical matching + reranking
& Vector + inverted indexes \\

Symphony~\cite{DBLP:conf/cidr/ChenG0FM023}
& MUL
& Source metadata / contents
& Cross-modal source vectors
& NL task / question
& Query--source semantic matching
& Vector index + Top-$k$ retrieval \\

PANGAEA-GPT~\cite{pantiukhin2026hierarchical}
& MUL
& Dataset metadata / descriptions
& Metadata records
& NL scientific request
& Iterative semantic relevance
& Repository search + Query refinement \\
\papersep
\hline

\rowcolor{gray!12}
\multicolumn{7}{l}{\textit{Downstream Utility-driven Dataset Search}} \\
\hline
\papersep

METAM~\cite{galhotra2023metam}
& TAB
& Data profiles / candidate relations / task feedback
& Profile vectors
& Dataset + Task
& Downstream utility
& Clustering + Adaptive evaluation \\

Saibot~\cite{huang2023saibot}
& TAB
& Candidate data / task statistics
& Privatized statistics
& Dataset + Task
& Model utility
& Precomputation + Factorization \\

Mileena~\cite{huang2024mileena}
& TAB
& Candidate data / task statistics
& Statistical sketches
& Dataset + Task
& Proxy task utility
& Sketching + Pruning \\
\papersep

Toast~\cite{zhu2025toast}
& SPA
& Candidate data / task feedback / cost
& Candidate augmentations + cost profiles
& Dataset + Task
& Downstream utility + cost
& Candidate search + Iterative selection \\

MODis~\cite{wang2025generating}
& TAB / GRA
& Data sources / model-performance measures
& Multi-objective states
& Datasets + Model + Objectives
& Multi-objective utility
& Bidirectional search \\

\hline

\end{tabular}%
}

\vspace{1pt}

\begin{minipage}{0.98\textwidth}
\tiny
\textit{Note.}
\textbf{Modality:}
MUL = Multiple modalities;
TAB = Tabular;
SPA = Spatial;
GRA = Graph.
\textbf{Representation:}
KG = Knowledge Graph.
\textbf{User Input:}
NL = Natural Language.
\end{minipage}

\end{table*}

\vspace{2.5pt}
\noindent\textbf{Task-aware Dataset Recommendation.}
Task-aware dataset recommendation takes a research or analytical need as the
user input and retrieves datasets suitable for carrying out the specified
task. The central difficulty is that the user input typically describes
\emph{what the user intends to accomplish}, whereas dataset records primarily
describe \emph{what the data contain}. Relevance estimation must therefore
bridge the gap between task requirements and dataset characteristics.

DataFinder~\cite{viswanathan2023datafinder} formulates task-aware dataset
recommendation as semantic retrieval from research needs to datasets. It
represents datasets using descriptions, structured metadata, and information
from related papers, and learns a bi-encoder to estimate relevance between a
research request and candidate datasets. KATS~\cite{wei2026kats} further
incorporates explicit task--dataset relations into retrieval. It constructs a
knowledge graph connecting scientific tasks and datasets, retrieves
semantically related task nodes from the user input, and propagates their
relevance through the graph to rank associated datasets. This allows dataset
relevance to reflect not only textual similarity but also observed
relationships between tasks and datasets.

Recent systems further extend task-aware recommendation through richer intent
modeling, domain knowledge, and heterogeneous data sources. ScienceDB
AI~\cite{long2026sciencedb} transforms multi-turn scientific requests into
structured search requirements and uses them to retrieve and rerank datasets
from a large scientific repository. Yu et
al.~\cite{yu2026bringing} focus on Earth-observation data discovery,
using task--dataset relations derived from a domain knowledge graph together
with neural and lexical retrieval to rank datasets relevant to
natural-language research needs. Symphony~\cite{DBLP:conf/cidr/ChenG0FM023}
retrieves relevant databases, standalone tables, and documents from a
multimodal data lake according to a natural-language question before
source-specific processing is performed. PANGAEA-GPT~\cite{pantiukhin2026hierarchical}
similarly searches large scientific archives for datasets relevant to a
natural-language scientific request and subsequently routes the retrieved
data according to their types. Overall, these methods extend task-aware
relevance estimation from direct semantic matching toward relation-aware,
domain-grounded, and heterogeneous-source dataset retrieval.

\vspace{2.5pt}
\noindent\textbf{Downstream Utility-driven Dataset Search.}
A second line of work defines dataset relevance through the contribution of
candidate data to a downstream task. The user input typically consists of a
base dataset together with an analytical or machine learning task, while
candidate datasets can be incorporated through join, union, or other
augmentation operations. Unlike dataset relationship search, where relevance
is determined by whether a desired inter-dataset relationship can be
established, utility-driven search further evaluates whether incorporating a
candidate improves the intended downstream use. For a base dataset $D$ and
candidate augmentation $D_i$, this principle can be expressed as
\[
\Delta U_T(D_i \mid D)
=
U_T(\operatorname{Augment}(D,D_i))
-
U_T(D),
\]
where $U_T$ denotes the utility of the resulting data for task $T$.

Specifically, METAM~\cite{galhotra2023metam} incorporates downstream-task
feedback directly into dataset discovery. It represents candidate
augmentations through data profiles, groups similar candidates, and
selectively evaluates promising datasets on the target task, using observed
utility to guide subsequent search. Saibot~\cite{huang2023saibot} and
Mileena~\cite{huang2024mileena} further reduce the cost of task-dependent
dataset evaluation. Both search join- or union-compatible datasets according
to their contribution to machine-learning performance, while using reusable
statistics or sketches to estimate candidate utility more efficiently under
privacy and latency constraints. Toast~\cite{zhu2025toast} extends
utility-driven search to spatio-temporal data by searching unionable
trajectory datasets and joinable contextual datasets and selecting
augmentations according to downstream benefit and execution cost.
MODis~\cite{wang2025generating} further generalizes dataset utility from a
single downstream objective to multiple model-performance measures,
selecting and integrating data sources according to Pareto dominance.
Together, these methods move downstream utility-driven dataset search toward
more efficient, cost-aware, privacy-aware, and multi-objective estimation of
task-dependent dataset utility.

\vspace{2.5pt}
\noindent\textbf{Discussion.}
Task-oriented dataset search differs from the preceding paradigms because
dataset relevance depends explicitly on the user's intended task. The same
candidate dataset may be suitable for one analytical need but less useful for
another, and its contribution may also depend on the base data with which it
is used. This task dependence creates a tension between the fidelity and cost
of relevance estimation. Task-aware recommendation often relies on semantic,
relational, or domain-specific evidence to infer suitability without
executing the target task, whereas utility-driven search can obtain a more
direct signal by integrating candidate data and evaluating downstream
performance, at substantially higher cost. Existing systems therefore use
learned relevance models, data profiles, reusable statistics, sketches, and
other proxy signals to reduce expensive task-specific evaluation. A central
challenge is to estimate task-dependent relevance faithfully and efficiently,
while preserving the distinctions among semantic suitability, relationship
compatibility, and actual downstream utility.

\end{sloppypar}

%% file: sections/sec5.tex
\begin{sloppypar}

\section{LLM for Dataset Search}
\label{sec:llm_for_dataset_search}

Trained on large and heterogeneous corpora, LLMs acquire broad semantic
knowledge, strong instruction-following capabilities, and the ability to
interpret and reason over diverse forms of information with limited
task-specific supervision. Accordingly, LLMs have been increasingly applied
to data management~\cite{li2024llmdata}, information
retrieval~\cite{breuer2025llmir}, and data discovery and
integration~\cite{DBLP:journals/debu/FreireFFKLPSSW25}, and have shown
potential at different stages of dataset search.

Following the analysis framework introduced in
Section~\ref{sec: Data Extraction and Analysis}, we examine LLM-enhanced
dataset search in terms of the \emph{LLM inputs}, \emph{LLM outputs}, and
\emph{how LLM-generated information is integrated into the dataset search
pipeline}. Based on the stage of integration, we organize existing methods
into four aspects. For \emph{search evidence construction}, LLMs can generate
semantic descriptions and construct repository-level relations that expose
information not directly available from raw
metadata~\cite{zhang2026autoddg,wei2026kats}. For \emph{dataset
representation and indexing}, they can generate retrieval supervision,
construct semantic views, and support identifier-based generative
indexing~\cite{silva2024improving,terrenzi2026piper,zhang2026generative}.
For \emph{user input and search planning}, LLMs can interpret, refine, and
decompose natural-language requests into explicit search requirements,
retrieval specifications, and executable search
plans~\cite{long2026sciencedb,sun2026multi,pantiukhin2026hierarchical,
li2026octopus}. For \emph{relevance estimation}, they can reason over
candidate contents, structures, and multiple modalities to validate or
rerank retrieved datasets~\cite{soliman2026coret,xu2026efficient,
yu2026bringing}. We focus on studies in which LLM-generated outputs are
explicitly used in at least one of these stages of the dataset search
process.

\subsection{LLM-enhanced Search Evidence Construction}

Dataset search relies on evidence from metadata, dataset contents, and
contextual sources to estimate relevance to user needs. LLMs provide a
flexible way to derive additional search evidence from datasets themselves
and from external contexts such as scientific publications. Existing studies
mainly use LLMs to enrich individual datasets with searchable semantic
evidence or to construct repository-level evidence that connects datasets
with categories, publications, tasks, and usage contexts. These generated
signals can complement incomplete or coarse-grained metadata during retrieval.
Table~\ref{tab:llm_search_evidence} summarizes representative methods.

\begin{table*}[t]
\centering
\caption{Summary of representative methods for LLM-enhanced search evidence construction.}
\label{tab:llm_search_evidence}

\fontsize{6.2pt}{7.0pt}\selectfont
\setlength{\tabcolsep}{4pt}
\renewcommand{\arraystretch}{0.90}

\resizebox{0.995\textwidth}{!}{%
\begin{tabular}{
>{\raggedright\arraybackslash\linespread{0.86}\selectfont}p{1.45cm}
>{\raggedright\arraybackslash\linespread{0.86}\selectfont}p{2.80cm}
>{\raggedright\arraybackslash\linespread{0.86}\selectfont}p{3.20cm}
>{\raggedright\arraybackslash\linespread{0.86}\selectfont}p{5.20cm}
}

\hline
\rowcolor{gray!30}
\textbf{Method} &
\textbf{LLM Input} &
\textbf{LLM Output} &
\textbf{Search Integration} \\
\hline


\rowcolor{gray!12}
\multicolumn{4}{l}{\textit{Dataset-level Evidence Generation}} \\
\hline
\papersep

AutoDDG~\cite{zhang2026autoddg}
& Column names / sample values / data profiles
& Semantic profiles / user- and search-focused descriptions
& Descriptions are indexed for retrieval; user-focused descriptions support relevance assessment
\\
\papersep

CN-PDS~\cite{chen2026cnpds}
& Heterogeneous metadata / dataset content chunks
& Unified metadata / enriched dataset descriptions
& Consolidated and enriched metadata are used for unified dataset retrieval
\\
\papersep

Terrenzi \& Ayvaz~\cite{terrenzi2026faithful}
& Original metadata / RDF profiles / RDF graph context
& Rewritten / synthetic dataset metadata
& Generated metadata are used as dataset surrogates for sparse and dense retrieval
\\
\papersep
\hline


\rowcolor{gray!12}
\multicolumn{4}{l}{\textit{Repository-level Evidence Construction}} \\
\hline
\papersep

LEDD~\cite{an2025ledd}
& Source / table / column facets / clustered nodes
& Semantic summaries / hierarchical category labels
& Generated semantics are organized into a hierarchical catalog for table and column search
\\
\papersep

ChatPD~\cite{xu2025chatpd}
& Paper title / abstract / dataset-related sections
& Structured dataset descriptions / metadata records
& Extracted records are resolved into a paper--dataset network for dataset query and recommendation
\\
\papersep

KATS~\cite{wei2026kats}
& Scientific papers / dataset mentions
& Task--dataset pairs / enriched task descriptions / keywords
& Extracted evidence is organized into a task--dataset knowledge graph for retrieval and ranking
\\
\papersep

Chenarides et al.~\cite{chenarides2026data}
& Publication text / matched dataset mentions
& Dataset-use judgments / usage context / related dataset names
& Generated usage evidence is aggregated into dataset-level descriptors for indexing and discovery
\\

\hline

\end{tabular}%
}

\end{table*}

\vspace{2.5pt}
\noindent\textbf{Dataset-level Evidence Generation.}
LLMs can derive richer semantic evidence for an individual dataset from its
metadata, contents, and profiles, producing searchable descriptions that
better expose what the dataset contains and how it may satisfy user needs.
AutoDDG~\cite{zhang2026autoddg} uses column names, sampled values, and data
profiles to generate semantic and search-oriented descriptions that are
subsequently indexed for dataset retrieval. CN-PDS~\cite{chen2026cnpds}
further consolidates heterogeneous metadata across public data portals and
enriches dataset descriptions with retrieved contents through multi-agent
processing and RAG, providing unified evidence for cross-source search.
Terrenzi and Ayvaz~\cite{terrenzi2026faithful} generate rewritten or
synthetic metadata from original metadata, dataset profiles, and RDF context,
and directly use the generated metadata as surrogates for sparse and dense
retrieval. Across these methods, LLMs make dataset-specific information more
explicit and directly usable as evidence for subsequent retrieval. Together,
these studies move from enriching existing metadata toward constructing
retrieval-oriented evidence explicitly designed to improve dataset
findability.

\vspace{2.5pt}
\noindent\textbf{Repository-level Evidence Construction.}
Beyond enriching individual datasets, LLMs can construct search evidence
from relationships among datasets and other entities, such as repository
categories, scientific publications, tasks, and usage contexts.
LEDD~\cite{an2025ledd} summarizes source-, table-, and column-level
information and organizes LLM-generated semantic labels into a hierarchical
catalog for data discovery. ChatPD~\cite{xu2025chatpd} extracts structured
dataset descriptions from scientific papers and resolves them into a
paper--dataset network for dataset query and recommendation.
KATS~\cite{wei2026kats} uses multiple LLM agents to identify task--dataset
pairs and enrich task descriptions, constructing a task--dataset knowledge
graph for retrieval and ranking. Chenarides et
al.~\cite{chenarides2026data} derive dataset-use evidence from publication
text and aggregate it into searchable descriptors that capture how datasets
are used in practice. Together, these methods extend search evidence beyond
the semantics of individual datasets by exposing repository structure and
contextual relationships among datasets, publications, tasks, and usage
scenarios.

\vspace{2.5pt}
\noindent\textbf{Discussion.}
LLM-enhanced search evidence construction enriches dataset descriptions with
semantic, relational, and usage-aware information, but current methods remain
sensitive to incomplete source context and generation errors. Once
incorporated into searchable descriptions, catalogs, or knowledge graphs,
such errors may propagate to subsequent retrieval and become difficult to
separate from evidence grounded in the original sources. Moreover, enriching
search evidence may improve findability at the cost of introducing
unsupported semantics. A central challenge is therefore to balance retrieval
effectiveness with grounding and faithfulness as LLM-generated evidence
becomes increasingly influential in dataset search.

\subsection{LLM-enhanced Dataset Representation and Indexing}

LLMs can interpret complex data context and generate semantic descriptions,
synthetic supervision, and structured outputs, making them useful for
constructing search-oriented dataset representations~\cite{li2024llmdata,breuer2025llmir}.
These capabilities have increasingly been exploited in dataset representation
and indexing, which determine how dataset information is encoded and accessed
during retrieval. Existing studies use LLMs to generate supervision for
representation learning, construct semantic views or relations that are
subsequently represented and indexed, and support generative retrieval through
mappings from queries to dataset identifiers.
Table~\ref{tab:llm_representation_indexing} summarizes representative methods.

\begin{table*}[t]
\centering
\caption{Summary of representative methods for LLM-enhanced dataset representation and indexing.}
\label{tab:llm_representation_indexing}

\fontsize{6.2pt}{7.0pt}\selectfont
\setlength{\tabcolsep}{4pt}
\renewcommand{\arraystretch}{0.90}

\resizebox{0.995\textwidth}{!}{%
\begin{tabular}{
>{\raggedright\arraybackslash\linespread{0.90}\selectfont}p{1.45cm}
>{\raggedright\arraybackslash\linespread{0.90}\selectfont}p{3.00cm}
>{\raggedright\arraybackslash\linespread{0.90}\selectfont}p{2.4cm}
>{\raggedright\arraybackslash\linespread{0.90}\selectfont}p{5.5cm}
}

\hline
\rowcolor{gray!30}
\textbf{Method} &
\textbf{LLM Input} &
\textbf{LLM Output} &
\textbf{Search Integration} \\
\hline


\rowcolor{gray!12}
\multicolumn{4}{l}{\textit{Retrieval Supervision for Representation Learning}} \\
\hline
\papersep

DAPDR~\cite{silva2024improving}
&
Dataset title / description / keywords
&
Synthetic dataset search queries
&
Synthetic query--dataset pairs are used to fine-tune dense retrievers for dataset ranking
\\
\papersep

Solo~\cite{wang2023solo}
&
Automatically sampled SQL queries
&
Natural-language questions
&
Generated question--table pairs provide repository-specific supervision for training the table retriever
\\
\papersep
\hline


\rowcolor{gray!12}
\multicolumn{4}{l}{\textit{Evidence Representation and Index Construction}} \\
\hline
\papersep

Pneuma~\cite{balaka2025pneuma}
&
Table columns / full schema context
&
Column descriptions / schema narrations
&
Schema narrations are combined with sampled rows and table context and stored in full-text and vector indexes
\\
\papersep

PIPER~\cite{terrenzi2026piper}
&
Full-table statistical profiles
&
User-oriented pseudoqueries
&
Pseudoqueries are independently embedded and stored as multiple semantic access paths in a vector database
\\
\papersep

HyperJoin~\cite{liu2026hyperjoin}
&
Join-key column / table and column names / sample rows
&
Semantically preserving column-name variants
&
Generated variants augment the join graph, whose relations are used to learn structure-aware column representations
\\
\papersep
\hline


\rowcolor{gray!12}
\multicolumn{4}{l}{\textit{Generative Indexing}} \\
\hline
\papersep

Birdie~\cite{guo2025birdie}
&
Sampled table content / table or query sequences
&
Synthetic queries / table-identifier sequences
&
Mappings from queries and tables to identifiers are encoded in model parameters for direct identifier-based retrieval
\\
\papersep

GenTUS~\cite{zhang2026generative}
&
Serialized query tables
&
Semantic-identifier sequences
&
Generated identifiers are constrained by a prefix trie and directly mapped to candidate tables without ANN lookup
\\

\hline

\end{tabular}%
}

\end{table*}

\vspace{2.5pt}
\noindent\textbf{Retrieval Supervision for Representation Learning.}
One role of language models in representation learning is to generate
otherwise scarce query--dataset supervision.
DAPDR~\cite{silva2024improving} generates synthetic search queries from
dataset titles, descriptions, and keywords and uses the resulting
query--dataset pairs to fine-tune dense retrievers.
Solo~\cite{wang2023solo} instead generates SQL queries from repository tables
and uses a fine-tuned T5 model to translate them into natural-language
questions, providing self-supervised question--table pairs for training its
table retriever. In both cases, the language model produces supervision
rather than the final indexed representation, shifting representation learning
from externally labeled queries toward repository-derived synthetic
supervision.

\vspace{2.5pt}
\noindent\textbf{Evidence Representation and Index Construction.}
Another line of work uses LLM-generated semantics as intermediate signals for
constructing retrieval-oriented representations and indexes.
Pneuma~\cite{balaka2025pneuma} combines LLM-generated column descriptions
with sampled rows and table context and stores the resulting representations
in full-text and vector indexes.
PIPER~\cite{terrenzi2026piper} generates multiple user-oriented pseudoqueries
from a full-table statistical profile and independently embeds them, providing
multiple semantic access paths to the same dataset in a vector database.
HyperJoin~\cite{liu2026hyperjoin} uses LLM-generated variants of join-key
names to augment a join graph, whose connected components are further
converted into inter-table hyperedges for learning structure-aware column
representations. These methods use LLM-generated semantics not merely as
additional searchable evidence, but as intermediate signals for constructing
multi-view, relation-aware, and indexable dataset representations.

\vspace{2.5pt}
\noindent\textbf{Generative Indexing.}
A more recent direction changes not only how datasets are represented, but
also how they are addressed during retrieval.
Birdie~\cite{guo2025birdie} assigns prefix-aware table identifiers and uses
an LLM-based query generator to create table-specific synthetic queries. An
encoder--decoder model then learns mappings from both tables and queries to
their identifiers, encoding part of the retrieval structure in model
parameters and using isolated LoRA memory units to incorporate newly indexed
tables.
GenTUS~\cite{zhang2026generative} extends generative retrieval to table union
search. Its semantic identifiers are constructed separately through a table
encoder and unionability-aware residual quantization, while a T5 generator
learns to produce identifiers of unionable tables directly from a serialized
query table. Trie-constrained decoding restricts generation to valid
identifiers and enables new tables to be inserted into the candidate space
without retraining the generator. Together, these methods shift dataset
retrieval from conventional nearest-neighbor lookup toward identifier-based
generative retrieval, while illustrating different ways of separating
identifier construction, repository storage, and generative search.

\vspace{2.5pt}
\noindent\textbf{Discussion.}
LLM-enhanced representation and indexing shape which aspects of a dataset are
readily accessible at retrieval time. Semantic narrations and pseudoqueries
can expose meanings that may be difficult to capture with a single vector,
while information omitted during offline representation construction may
remain difficult to recover during later retrieval. Multi-view and
relation-aware representations can improve semantic coverage at the cost of
larger or more complex indexes. Generative indexing further shifts part of
repository addressing from external vector lookup toward learned identifiers
and model-based mappings, making identifier consistency, constrained
generation, and incremental updates important concerns. A key challenge is
therefore to balance semantic coverage with compact, stable, and updateable
representations as dataset indexing becomes increasingly mediated by
language models.

\subsection{LLM-enhanced User Input and Search Planning}
\label{sec:llm_user_input_planning}

Semantic and interactive search over data repositories has been explored in
earlier systems such as Olio~\cite{setlur2023olio}. Building on this direction,
LLMs can further interpret underspecified natural-language requests and
generate structured search requirements that make implicit needs, constraints,
and preferences more explicit before retrieval, thereby reducing ambiguity in
how the original information need is translated into search operations.
This makes them suitable for user-input processing and search
planning~\cite{breuer2025llmir,long2026sciencedb}. Existing methods mainly
support three roles: interpreting and refining user needs into explicit intents
or constraints, translating them into retrieval-oriented specifications such as
schema elements or subqueries, and planning how these requirements should be
satisfied through appropriate queries, sources, tools, or search operations.
These functions span the process from understanding \emph{what to search for},
through specifying \emph{what should be retrieved}, to deciding
\emph{how to search}. Table~\ref{tab:llm_user_input_planning} summarizes
representative methods.

\begin{table*}[t]
\centering
\caption{Summary of representative methods for LLM-enhanced user input and search planning.}
\label{tab:llm_user_input_planning}

\fontsize{6.2pt}{7.0pt}\selectfont
\setlength{\tabcolsep}{4.0pt}
\renewcommand{\arraystretch}{0.90}

\resizebox{0.995\textwidth}{!}{%
\begin{tabular}{
>{\raggedright\arraybackslash\linespread{0.90}\selectfont}p{1.85cm}
>{\raggedright\arraybackslash\linespread{0.90}\selectfont}p{3.15cm}
>{\raggedright\arraybackslash\linespread{0.90}\selectfont}p{3.45cm}
>{\raggedright\arraybackslash\linespread{0.90}\selectfont}p{5.00cm}
}

\hline
\rowcolor{gray!30}
\textbf{Method} &
\textbf{LLM Input} &
\textbf{LLM Output} &
\textbf{Search Integration} \\
\hline


\rowcolor{gray!12}
\multicolumn{4}{l}{\textit{Query Understanding and Interactive Reformulation}} \\
\hline
\papersep

ScienceDB AI~\cite{long2026sciencedb}
&
User query / dialogue history
&
Structured scientific intent / experimental requirements
&
Guides dataset retrieval, filtering, and conversational refinement
\\
\papersep

SemExplorer~\cite{wei2025semexplorer}
&
Natural-language query
&
Semantic query representation / statistical filters
&
Supports semantic matching and explicit constraint-based filtering
\\
\papersep

Wetzel et al.~\cite{wetzel2026query}
&
User query / conversation context
&
Search-readiness decision / clarification question / subqueries / spatio-temporal constraints
&
Controls retrieval initiation and parameterizes federated dataset search
\\
\papersep

DataScout~\cite{lin2025datascout}
&
User query / retrieved-result themes and attributes
&
Hypothetical target schemas / query reformulations / filter concepts
&
Supports initial retrieval and result-grounded iterative refinement
\\
\papersep

Pneuma-Seeker~\cite{balaka2026pneumaseeker}
&
User query / feedback / retrieved evidence
&
Target relational schema / schema revisions
&
Uses the evolving schema to guide dataset discovery and preparation
\\
\papersep

PublicAgent~\cite{publicagent2026}
&
User query / clarification responses
&
Clarification questions / enhanced query
&
Uses the enhanced query to drive subsequent repository search
\\

\hline


\rowcolor{gray!12}
\multicolumn{4}{l}{\textit{Query Decomposition and Retrieval Specification}} \\
\hline
\papersep

Octopus~\cite{li2026octopus}
&
Natural-language query
&
Column mentions / SQL sketch
&
Provides schema- and value-level cues for subsequent table and content retrieval
\\
\papersep

FGTR~\cite{sun2026fgtr}
&
User query / semantized table schemas
&
Key query elements / relevant columns / value-range constraints
&
Specifies schema- and value-level conditions for hierarchical retrieval
\\
\papersep

DMRAL~\cite{luo2026decomposition}
&
User question / table-structure evidence / coverage gaps
&
Table-aligned subquestions / residual subquestions
&
Guides coverage-aware retrieval and complementary-table recovery
\\
\papersep
\hline


\rowcolor{gray!12}
\multicolumn{4}{l}{\textit{Search Planning and Source Orchestration}} \\
\hline
\papersep

Q2DataAgent~\cite{yang2025quest2dataagent}
&
Natural-language research question
&
Structured dataset-demand subtasks
&
Routes dataset-demand subtasks to appropriate retrieval strategies
\\
\papersep

LakeAgents~\cite{tang2026lakeagents}
&
User request / task and table context
&
Structured search parameters / agent decisions
&
Coordinates table selection and join/augmentation-column discovery
\\
\papersep

Metadata Reasoner~\cite{zhang2026metadatareasoning}
&
Analytical task / retrieved metadata / tool observations
&
Search constraints / search plans / tool calls
&
Controls iterative table search, metadata checks, and strategy revision
\\
\papersep

PANGAEA-GPT~\cite{pantiukhin2026hierarchical}
&
Natural-language query / retrieved metadata
&
Metadata constraints / Boolean queries / refined queries
&
Drives iterative scientific-archive search and query refinement
\\
\papersep

Terrenzi et al.~\cite{terrenzi2026agentic}
&
User query / hybrid-retrieval results
&
Planned or reformulated queries / result-sufficiency decisions
&
Controls repeated hybrid retrieval and query reformulation
\\
\papersep

Blackboard System~\cite{salemi2025blackboard}
&
Information need / retrieved evidence
&
Posted subtasks / agent responses / updated search requests
&
Coordinates distributed retrieval and follow-up search through shared state
\\

\hline

\end{tabular}%
}

\end{table*}

\vspace{2.5pt}
\noindent\textbf{Query Understanding and Interactive Reformulation.}
LLMs can transform underspecified natural-language requests into explicit search requirements and interactively refine them before and during retrieval. ScienceDB AI~\cite{long2026sciencedb} parses a user query together with dialogue history into structured scientific intent, including the research subject and task, required data modality, experimental settings, and evaluation criteria, which are subsequently used for dataset retrieval and filtering. SemExplorer~\cite{wei2025semexplorer} interprets natural-language requests into semantic query representations and statistical filters, allowing thematic similarity and explicit data conditions to jointly constrain dataset search. Wetzel et al.~\cite{wetzel2026query} further determine whether thematic and spatio-temporal requirements are sufficiently specified. When information is missing, the LLM generates clarification questions and incorporates the user's responses before retrieval; otherwise, the request is decomposed into search queries together with spatial and temporal constraints for federated retrieval. PublicAgent~\cite{publicagent2026} similarly supports interactive ambiguity resolution by identifying unclear terms and missing constraints, asking targeted clarification questions, and consolidating the responses into an enhanced query for subsequent dataset discovery.

Interaction can also continue after initial retrieval, allowing retrieved datasets to inform subsequent reformulation. DataScout~\cite{lin2025datascout} generates multiple hypothetical target schemas from a user query and uses them as semantic surrogates for initial retrieval. It then analyzes themes and attributes in the returned datasets and generates result-grounded query reformulations and filter concepts, allowing subsequent searches to be progressively refined according to the available data space. Pneuma-Seeker~\cite{balaka2026pneumaseeker} makes the evolving information need explicit through a target relational schema describing the attributes, types, and semantics expected in the desired result. The schema can be inspected and iteratively revised according to user feedback and retrieved evidence, and the updated specification subsequently guides dataset discovery and preparation. Together, these methods move user-input processing from one-shot intent extraction toward interactive clarification, result-grounded reformulation, and inspectable representations of evolving information needs.

\vspace{2.5pt}
\noindent\textbf{Query Decomposition and Retrieval Specification.}
Beyond understanding or reformulating a user's dataset need, LLMs can translate it into finer-grained specifications that downstream retrieval components can directly use. Octopus~\cite{li2026octopus} infers column mentions and a SQL sketch from a natural-language query, providing schema- and value-level cues for subsequent table and content retrieval. FGTR~\cite{sun2026fgtr} extracts key query elements, maps them to relevant columns in semantized table schemas, and derives value-range constraints for hierarchical schema- and cell-level retrieval. DMRAL~\cite{luo2026decomposition} decomposes complex questions into table-aligned subquestions grounded in repository structure and generates residual subquestions when the retrieved tables do not fully cover the original information need. These methods extend query processing from high-level intent interpretation toward retrieval-oriented decomposition and specification at the schema, value, and subquestion levels.

\vspace{2.5pt}
\noindent\textbf{Search Planning and Source Orchestration.}
Once a dataset need has been translated into searchable requirements, LLMs can determine how those requirements should be satisfied through available retrieval mechanisms. Quest2DataAgent~\cite{yang2025quest2dataagent} decomposes a natural-language research question into structured dataset-demand subtasks and routes them to knowledge-graph, semantic, or rule-based retrieval strategies. LakeAgents~\cite{tang2026lakeagents} organizes table augmentation through specialized agents for table selection, join-column identification, and augmentation-column selection. These methods translate high-level dataset needs into structured subtasks and coordinated sequences of specialized retrieval actions.

More adaptive methods allow intermediate evidence to influence subsequent search actions. Metadata Reasoner~\cite{zhang2026metadatareasoning} constructs search constraints and plans from an analytical task and iteratively alternates semantic table search with metadata, value, and joinability checks to revise its strategy. PANGAEA-GPT~\cite{pantiukhin2026hierarchical} generates metadata constraints and Boolean queries for scientific archives and refines subsequent searches according to returned metadata. Terrenzi et al.~\cite{terrenzi2026agentic} similarly use retrieval results to reformulate queries and decide whether further hybrid retrieval is required. The Blackboard System~\cite{salemi2025blackboard} further decentralizes orchestration by maintaining the current information need and retrieved evidence in a shared blackboard, through which specialized retrieval agents contribute and trigger subsequent searches without explicit assignment of every action by a central planner. Together, these methods extend search planning from structured task decomposition toward adaptive and decentralized orchestration driven by intermediate retrieval evidence.

\vspace{2.5pt}
\noindent\textbf{Discussion.}
LLM-enhanced user input and search planning make complex dataset needs easier to operationalize, but current methods remain sensitive to errors introduced during interpretation and planning. Automatically inferred constraints, schemas, or subqueries may cause query drift or prematurely restrict the search space, especially when underspecified user needs are converted into fine-grained retrieval specifications. Such errors can further propagate to downstream retrieval, where incorrect schema mappings, value constraints, or decompositions may prevent relevant datasets from being considered. Interactive clarification and result-grounded refinement can alleviate some of these issues, but excessive interaction increases user effort, while iterative or agentic planning may introduce inconsistencies with the original intent as search decisions accumulate. A key challenge is therefore to balance search autonomy with intent fidelity while keeping intermediate requirements, retrieval specifications, and planning decisions transparent enough for users to inspect and correct.

\subsection{LLM-enhanced Relevance Estimation}
\label{sec:llm_relevance_estimation}

LLMs can reason jointly over user requirements and heterogeneous candidate
evidence, including content, structural, and contextual signals, making them
suitable for context-sensitive relevance estimation beyond fixed similarity
functions~\cite{soliman2026coret,xu2026efficient,yu2026bringing}. Existing
methods use LLMs either to directly assess candidate relationships or to refine
candidates returned by scalable first-stage retrieval, including candidate
validation, subset selection, and reranking. In both cases, the final relevance
decision can incorporate richer candidate context than a single predefined
similarity score. Table~\ref{tab:llm_relevance_estimation} summarizes
representative methods.

\begin{table*}[t]
\centering
\caption{Summary of representative methods for LLM-enhanced relevance estimation.}
\label{tab:llm_relevance_estimation}

\fontsize{6.2pt}{7.0pt}\selectfont
\setlength{\tabcolsep}{2.5pt}
\renewcommand{\arraystretch}{0.90}

\resizebox{0.995\textwidth}{!}{%
\begin{tabular}{
>{\raggedright\arraybackslash\linespread{0.90}\selectfont}p{1.75cm}
>{\raggedright\arraybackslash\linespread{0.90}\selectfont}p{3.15cm}
>{\raggedright\arraybackslash\linespread{0.90}\selectfont}p{3.45cm}
>{\raggedright\arraybackslash\linespread{0.90}\selectfont}p{5.00cm}
}

\hline
\rowcolor{gray!30}
\textbf{Method} &
\textbf{LLM Input} &
\textbf{LLM Output} &
\textbf{Search Integration} \\
\hline
\papersep
DiscoverGPT~\cite{hu2025discovergpt}
&
Serialized table / column pairs
&
Column-relatedness decisions
&
Estimates semantic joinability for related-table discovery
\\
\papersep

Pneuma~\cite{balaka2025pneuma}
&
User question / hybrid-retrieved candidate documents
&
Binary relevance judgments
&
Filters and reorders candidates produced by combined full-text and vector retrieval
\\
\papersep

CN-PDS~\cite{chen2026cnpds}
&
User query / BM25-retrieved datasets / retrieved content evidence
&
Reranked dataset list
&
Reranks first-stage BM25 results using LLM-based relevance assessment
\\
\papersep

Xiao et al.~\cite{xiao2026unified}
&
Discovery task / filtered candidate-table context
&
Contextual join / union judgments
&
Validates candidates after metadata filtering and vector search
\\
\papersep

CORE-T~\cite{soliman2026coret}
&
Query / top-$k$ table previews / compatibility evidence
&
Coherent selected table subset
&
Prunes dense candidates to a compact and joinable table set
\\
\papersep

TabRAG~\cite{xu2026efficient}
&
Natural-language query / candidate table images
&
Pointwise / listwise relevance judgments
&
Reranks visual table candidates before downstream reasoning
\\
\papersep

TabRank~\cite{singh2026tabrank}
&
Query / candidate table list
&
Reasoning traces / ranking labels
&
Distills reasoning into compact listwise table rerankers
\\
\papersep

Yu et al.~\cite{yu2026bringing}
&
Scientific query / top-$k$ dataset metadata / external evidence
&
Ranked candidate identifiers
&
Reorders top-$k$ datasets using tool-grounded relevance reasoning
\\

\hline

\end{tabular}%
}

\end{table*}

DiscoverGPT~\cite{hu2025discovergpt} fine-tunes an LLM with multiple
table-related auxiliary tasks and applies it to serialized table--column pairs
to predict column relatedness, providing semantic joinability judgments for
related-table discovery. Other systems apply LLM reasoning to candidates
produced by lightweight first-stage retrieval.
Pneuma~\cite{balaka2025pneuma} combines lexical and dense retrieval with
LLM-based candidate judgment, while CN-PDS~\cite{chen2026cnpds} applies
RAG-grounded LLM reranking to candidates retrieved from heterogeneous metadata
sources. Xiao et al.~\cite{xiao2026unified} further combine metadata filtering,
vector search, and LLM-based contextual reasoning in a cascaded framework,
where the final LLM layer validates joinable or unionable candidates.
CORE-T~\cite{soliman2026coret} moves beyond independent candidate judgments by
jointly considering the query, retrieved tables, and compatibility evidence to
select a compact and coherent table subset, allowing compatibility among the
selected tables to influence the final selection. Together, these methods
extend relevance estimation from direct pairwise judgments toward
retrieval-grounded, context-aware, and set-coherent candidate selection.

Recent methods further enrich the evidence and reasoning used during reranking.
TabRAG~\cite{xu2026efficient} first retrieves table images with a
visual--textual bi-encoder and then applies a multimodal LLM for pointwise or
listwise reranking, allowing both table content and visual layout to influence
relevance judgments and enabling candidates to be compared using evidence
beyond their textual representation. TabRank~\cite{singh2026tabrank} uses a
large reasoning model to generate reasoning traces and ranking supervision over
query--candidate lists and distills this information into compact listwise
table rerankers, introducing explicit reasoning supervision into
structured-data ranking. For scientific dataset discovery, Yu
et al.~\cite{yu2026bringing} apply LLM reasoning after first-stage retrieval
and further use an agentic reranker that can consult external evidence before
producing the final candidate ordering, allowing information beyond the
retrieved metadata to contribute to reranking. Overall, LLM-enhanced relevance
estimation is progressing from direct pairwise judgments toward
retrieval-grounded, set-aware, multimodal, listwise, and externally grounded
reasoning over dataset candidates.

\vspace{2.5pt}
\noindent\textbf{Discussion.}
LLM-enhanced relevance estimation extends dataset matching beyond fixed similarity functions by enabling context-sensitive reasoning over candidate contents, structures, sets, visual layouts, and external evidence. However, many existing methods operate on candidates surfaced by a lightweight first-stage retriever and therefore depend on the coverage of that candidate pool: relevant datasets omitted from the pool cannot be recovered by subsequent LLM reasoning. Expanding the candidate pool can improve coverage, but also increases the amount of candidate evidence to process and the computational cost of pointwise, listwise, multimodal, or agentic relevance assessment; for set-aware and listwise methods, it may also increase context requirements. Moreover, richer reasoning does not necessarily guarantee more reliable ranking, since LLM judgments may remain sensitive to candidate presentation, contextual evidence, and reasoning errors. A central challenge is therefore to balance candidate coverage, reasoning depth, and judgment reliability while keeping LLM-based relevance estimation computationally practical.

\section{Benchmarks for Dataset Search}
\label{sec:benchmarks}

This section reviews representative benchmarks for dataset search based on
the analysis dimensions introduced in Section~\ref{sec: Data Extraction and Analysis}.
Table~\ref{tab:dataset_search_benchmarks} summarizes their main characteristics,
while the following discussion examines benchmark collection and workload
construction, ground-truth construction, and evaluation metrics and protocols.

\begin{table*}[t]
\centering
\caption{Representative benchmarks for dataset search.}
\label{tab:dataset_search_benchmarks}
\footnotesize
\setlength{\tabcolsep}{3.2pt}
\renewcommand{\arraystretch}{1.06}

\resizebox{0.995\textwidth}{!}{%
\begin{tabular}{
>{\raggedright\arraybackslash}p{2.00cm}
>{\raggedright\arraybackslash}p{5.10cm}
>{\centering\arraybackslash}p{1.20cm}
>{\raggedright\arraybackslash}p{2.00cm}
>{\centering\arraybackslash}p{1.20cm}
>{\raggedright\arraybackslash}p{4.10cm}
}
\hline
\rowcolor{gray!30}
\textbf{Benchmark} &
\textbf{Benchmark Scale} &
\textbf{Dataset Modality} &
\textbf{User Input} &
\textbf{Evaluation Target} &
\textbf{Ground Truth} \\
\hline
\papersep

Balaka et al.~\cite{balaka2025pneuma} &
5 corpora; 1K content + 1K context queries/corpus;
78--10,330 tables/corpus &
TAB &
NL query &
CA &
Rule-based labeling; model annotation \\
\papersep

FollowTable~\cite{jin2026followtable} &
4 corpora; 60--300 queries/corpus;
267--1,386 instructions/corpus;
9,546--23,784 tables/corpus &
TAB &
NL query + instructions &
CA &
Model annotation; human validation \\
\papersep

TaR-ViR~\cite{li2026tarvir} &
7,377 train queries; 1,249 test queries;
81,839 table images &
TAB &
NL query &
CA &
Label reuse; model annotation; human validation \\
\papersep

SANTOS~\cite{khatiwada2023santos} &
2 corpora; 50--80 queries/corpus;
550--11,090 tables/corpus &
TAB &
Query table + column &
DR &
Synthetic construction; human annotation \\
\papersep

LakeBench~\cite{deng2024lakebench} &
2 corpora; 4,762--7,428 join queries/corpus;
4,580--6,812 union queries/corpus;
64,698--16,670,064 tables/corpus &
TAB &
Query table + column &
DR &
Synthetic construction; human annotation \\
\papersep

nlcTables~\cite{cui2025nlctables} &
647 queries; 22,080 tables &
TAB &
NL conditions + table &
CA; DR &
Synthetic construction; rule-based labeling;
human validation \\
\papersep

STSD~\cite{leventidis2024stsd} &
2 snapshots; 2,560--9,296 queries/snapshot;
238,038--457,714 tables/snapshot &
TAB &
Query table &
CA &
Rule-based labeling; human validation \\
\papersep

DataFinder-E~\cite{he2025gradase} &
2,523 queries; 7,650 datasets &
MUL &
Text query + exemplar datasets &
CA; TO &
Label reuse \\
\papersep

DataFinder~\cite{viswanathan2023datafinder} &
17,495 train queries; 392 test queries;
$>7{,}000$ datasets &
MUL &
Keywords / task description &
TO &
Usage-based labeling; human validation \\
\papersep

CS-TDS~\cite{wei2026kats} &
2 corpora; 47--204 task queries/corpus;
$\approx 1{,}217$--$4{,}180$ datasets/corpus &
MUL &
Task description &
TO &
Usage-based labeling; human validation \\
\papersep

YADL~\cite{cappuzzo2025retrieve} &
6 prediction tasks; 4 lake variants;
70--47,223 tables/variant &
TAB &
Base table + prediction target &
TO &
Task labels; utility evaluation \\
\papersep

Lu et al.~\cite{lu2023statcan} &
6,662 queries; 5,907 tables &
TAB &
Dialogue context &
TO &
Interaction-derived labels \\
\papersep

LakeQA~\cite{wang2026lakeqa} &
1,007 tasks; $\approx 40$M files &
TAB; DOC &
NL question &
TO &
Human annotation \\
\papersep

KramaBench~\cite{lai2026kramabench} &
104 tasks; 633 subtasks; 1,764 files &
TAB; DOC &
Task description &
TO &
Human annotation \\

\bottomrule
\end{tabular}%
}

\vspace{2pt}
\begin{minipage}{0.995\textwidth}
\scriptsize
\textit{Note:}
\textbf{Dataset Modality:}
TAB = tabular data;
DOC = textual data;
MUL = multiple modalities.
\textbf{User Input:}
NL = natural language.
\textbf{Evaluation Target:}
CA = content-aware search;
DR = dataset relationship search;
TO = task-oriented search.
\end{minipage}
\end{table*}

\vspace{2.5pt}
\noindent\textbf{Dataset Collection and Construction.}
Benchmark collections are commonly assembled from existing repositories and
processed to support specific dataset search settings. Open data portals and
web tables provide heterogeneous schemas and values for discovering joinable
or unionable tables~\cite{deng2024lakebench}, while Wikipedia tables support
semantic dataset search using their contents and associated
entities~\cite{leventidis2024stsd}. Scientific dataset catalogs and datasets
identified in research papers provide candidate collections for task-oriented
dataset search~\cite{viswanathan2023datafinder,wei2026kats}. Existing
resources can also be transformed or synthesized to expose particular search
challenges. TaR-ViR~\cite{li2026tarvir} represents tables as images to
preserve visual structure during retrieval. YADL~\cite{cappuzzo2025retrieve}
constructs tables by recombining information from a knowledge base, providing
controlled data lake configurations for dataset search and predictive data
augmentation.

\vspace{2.5pt}
\noindent\textbf{Query and Task Construction.}
Benchmark workloads are obtained from user interactions, source datasets, or
descriptions of downstream needs. Real dialogues preserve contextual
information about how users express and refine their requests, as in the
conversational retrieval benchmark of Lu et al.~\cite{lu2023statcan}.
Queries can also be generated from dataset contents. Balaka et
al.~\cite{balaka2025pneuma} use SQL as an intermediate representation for
generating content questions and construct additional questions from
synthetic dataset contexts. More specific workloads combine queries with
instructions, natural-language conditions, or relevant dataset
examples~\cite{jin2026followtable,cui2025nlctables,he2025gradase}. These
constructions test whether search systems can satisfy requirements beyond
broad topical relevance. For search within analytical workflows, questions
and tasks may require locating and combining information across multiple
files. LakeQA~\cite{wang2026lakeqa} constructs questions through chains of
document-grounded subquestions, while KramaBench~\cite{lai2026kramabench}
derives analytical tasks from published studies and reports.

\vspace{2.5pt}
\noindent\textbf{Ground-Truth Construction.}
Ground truth may be established during benchmark construction or annotated
after queries and candidate datasets have been collected. Splitting source
tables creates known relationships between queries and candidates, supporting
controlled relevance labels for dataset relationship
search~\cite{khatiwada2023santos}. Rule-based labels can additionally encode
explicit conditions or approximate semantic relevance through shared
categories and links~\cite{cui2025nlctables,leventidis2024stsd}.
Usage-based and interaction-derived labels identify datasets used in source
papers or returned through human interactions~\cite{viswanathan2023datafinder,
wei2026kats,lu2023statcan}, although these observed choices may omit
other suitable candidates. Human annotation and model-assisted judgments
extend coverage~\cite{jin2026followtable,li2026tarvir}, often through
candidate pooling, consistency checks, and manual
validation~\cite{deng2024lakebench,balaka2025pneuma}. Such procedures do not
necessarily produce exhaustive relevance judgments over the entire
collection. For downstream tasks, reference answers, executable solutions,
or predictive labels provide a basis for assessing whether retrieved data
support successful use~\cite{wang2026lakeqa,lai2026kramabench,
cappuzzo2025retrieve}. These references evaluate broader outcomes whose
quality also depends on subsequent processing.

\vspace{2.5pt}
\noindent\textbf{Evaluation Metrics and Protocols.}
Evaluation depends on what a benchmark treats as the target of dataset
search. Retrieval-oriented benchmarks assess how effectively relevant
datasets are identified or ranked within a candidate collection, whereas
workflow-oriented benchmarks may additionally evaluate whether the retrieved
data support successful downstream analysis or prediction. These two levels
of evaluation should be distinguished because downstream outcomes depend not
only on retrieval quality, but also on subsequent integration, processing,
and reasoning. Evaluation protocols also differ in how candidate sets and
relevance judgments are constructed, making the coverage and reliability of
ground truth important when interpreting results across benchmarks.

\vspace{2.5pt}
\noindent\textbf{Discussion.}
The reviewed benchmarks broaden evaluation from finding relevant datasets to
satisfying explicit requirements and supporting their subsequent use.
However, many benchmarks focus on tabular data, while collections
spanning multiple modalities are often searched primarily through textual
descriptions or metadata. Despite substantial research on spatial,
structure-aware, and cross-modal dataset search, these settings remain
underrepresented in the reviewed benchmarks. More comprehensive benchmarks
should cover a broader range of dataset modalities and relevance criteria,
incorporate realistic user needs, and provide reliable relevance judgments.
For downstream workflows, retrieval quality and task outcomes should be
evaluated separately to clarify the contribution of dataset search.

\end{sloppypar}

%% file: sections/sec6.tex
\begin{sloppypar}

\section{Future Directions and Open Issues}
\label{sec: Future Directions and Open Issues}

This section discusses emerging research directions and open issues in dataset
search. Each direction builds on existing progress reviewed in the preceding
sections and highlights limitations that remain insufficiently addressed.

\subsection{Generative Dataset Search}

Most existing dataset search methods follow a retrieval-style paradigm, in
which user inputs are matched against externally indexed metadata, content
representations, or modality-specific features, and candidate datasets are
then ranked according to the resulting relevance signals. Generative retrieval
offers an alternative formulation by directly generating corpus-grounded
targets rather than relying solely on matching against an external index,
such as entity names, textual substrings, document identifiers, or learned
semantic identifiers~\cite{li2025matching,metzler2021rethinking}.
Representative approaches include name- or substring-based
generation~\cite{cao2021autoregressive,bevilacqua2022autoregressive},
numerical or hierarchical document
identifiers~\cite{tay2022transformer,wang2022neural}, learned semantic
identifiers~\cite{sun2023learning,jin2024language}, and multi-view or
set-based identifiers~\cite{li2023multiview,zhang2024generative}.

As discussed in Section~\ref{sec:llm_for_dataset_search}, recent studies have
begun to apply generative retrieval to dataset search. Birdie~\cite{guo2025birdie}
learns mappings from tables and queries to table identifiers, while
GenTUS~\cite{zhang2026generative} generates semantic identifiers for
unionable tables through constrained decoding. These studies demonstrate the
feasibility of identifier-based dataset retrieval, but current work remains
concentrated on tabular settings and specific search objectives. Meanwhile,
broader generative retrieval research still focuses primarily on entities,
documents, passages, and individual multimodal
objects~\cite{sun2025zerogr,long2024generative,li2024generative,
zhang2024irgen}. Extending generative retrieval to complete datasets with
heterogeneous contents, metadata, internal structures, provenance, versions,
and inter-dataset relationships, while preserving these dataset-level
characteristics during retrieval, therefore remains an open research
direction. Several issues require further investigation:

\begin{itemize}

    \item \textbf{Dataset-level Identifiers and Structure Preservation.}
    Existing generative identifiers are mainly designed for relatively
    self-contained retrieval objects. Dataset-level identifiers need to
    preserve or provide access to information such as metadata, schemas,
    content summaries, modality, provenance, and potentially relationships
    such as joinability or unionability. Multi-view, lexical, hierarchical,
    and learned semantic identifiers provide useful starting
    points~\cite{li2023multiview,sun2023learning,wang2023novo,
    cheng2025descriptive}, but new designs are needed to represent
    dataset-specific structure without reducing heterogeneous datasets to
    flat textual surrogates.

    \item \textbf{Scalable Search and Dynamic Updating.}
    Generative retrieval commonly relies on constrained autoregressive
    decoding, whose cost and early-pruning errors may become more pronounced
    as the target space grows~\cite{zeng2024scalable,wu2025constrained,
    kim2026d3}. Hybrid generative--retrieval architectures provide one
    possible direction by combining generated candidates or identifiers with
    conventional indexes for verification or
    reranking~\cite{zhang2023model,yuan2024generative}. Dataset repositories
    are also dynamic, requiring efficient insertion, deletion, and version
    updates without repeatedly retraining the entire
    model~\cite{mehta2023dsi++,chen2023continual,
    zhang2026model}.

    \item \textbf{Reliability and Evaluation.}
    Generative dataset search may produce invalid identifiers, confuse
    structurally similar datasets, or retrieve outdated versions.
    Corpus-constrained decoding and external verification mechanisms can help
    ground generated results~\cite{bevilacqua2022autoregressive,
    wang2022neural,lee2023nonparametric}. Future benchmarks should evaluate
    not only ranking effectiveness, but also identifier validity, decoding
    cost, update efficiency, structural relevance, and downstream dataset
    utility. Comparisons with retrieval-style methods are particularly
    important because much of the current evidence for generative retrieval
    still comes from document-oriented
    collections~\cite{zhang2025replication,wu2025constrained}.
\end{itemize}

\subsection{Privacy-preserving and Federated Dataset Search}

As reviewed in Sections~\ref{sec:ds} and
\ref{sec:llm_for_dataset_search}, recent studies have begun to address
dataset search under privacy-constrained and distributed settings, including
privacy-preserving similarity and task-oriented search as well as federated
discovery across multiple repositories. However, these efforts remain
limited compared with the diversity of dataset modalities and relevance
notions considered in current dataset search. Many existing methods still
assume that candidate datasets, or at least their searchable representations,
can be centrally accessed and indexed by the search system. In practice,
potentially useful datasets may be private~\cite{huang2023saibot},
access-controlled~\cite{alper2024dspack}, or distributed across independent
organizations and repositories~\cite{chapman2020dataset,wetzel2026query}.
Data owners may also be unwilling or unable to expose raw dataset contents
under applicable privacy, governance, or access constraints. These settings
create a practical need for privacy-preserving and federated dataset search,
which aims to support dataset discovery and relevance estimation across
protected or distributed sources without requiring all datasets to be
centralized or their sensitive contents to be directly exposed.

Privacy-preserving search~\cite{sharma2022searchable} and federated
search~\cite{garba2023federated} have been extensively studied in related
fields. In dataset search, existing privacy-preserving methods mainly address
specific settings, including differentially private task-based table
augmentation~\cite{huang2023saibot} and encrypted similarity or relationship
search over spatial datasets~\cite{li2024privacy,zhang2025privacy}.
Federated dataset search has only recently begun to receive attention, with
existing studies addressing metadata-based discovery across scientific
repositories~\cite{wetzel2026query} and joinable or unionable table search
across decentralized data repositories~\cite{behme2026fedaugment}.
Extending these techniques to richer dataset modalities, relevance functions,
and heterogeneous distributed repositories raises several open challenges:

\begin{itemize}

    \item \textbf{Privacy-preserving Relevance Estimation.}
    Modern dataset search increasingly relies on fine-grained dataset
    contents, distributions, relationships, and downstream utility to
    estimate relevance. Protecting such information makes expressive
    relevance estimation difficult because useful relevance signals must be
    preserved while sensitive dataset and query information remain
    hidden~\cite{zhu2025compass}. The challenge is therefore to support richer
    dataset-level relevance functions over protected data without
    substantially degrading retrieval effectiveness or scalability.

    \item \textbf{Federated Search across Heterogeneous Repositories.}
    Independently managed repositories may differ in metadata schemas, query
    interfaces, representations, retrieval models, and supported search
    operations. Candidates and relevance scores returned by different sources
    may therefore not be directly comparable~\cite{garba2025snippet}.
    Effective federated dataset search must support source selection, query
    adaptation, and result integration under such heterogeneity.

    \item \textbf{Privacy--Effectiveness--Efficiency Trade-off.}
    Stronger privacy protection can restrict the information or operations
    available for relevance estimation while introducing additional
    computation and communication overhead~\cite{soleimani2025weave}.
    These costs may be further amplified when search spans multiple
    distributed data owners. Balancing meaningful privacy guarantees with
    retrieval effectiveness and search efficiency therefore remains a
    fundamental challenge.

\end{itemize}

\subsection{Dataset Quality, Provenance, and Trust}

As reviewed in Sections~\ref{sec: systems}--\ref{sec:llm_for_dataset_search},
existing dataset search systems and methods primarily emphasize dataset
discoverability and relevance, while quality, provenance, and trust are less
directly incorporated into retrieval and ranking. However, relevance alone
does not guarantee that a retrieved dataset is suitable for reuse. Datasets
with similar topical or structural relevance may differ substantially in
completeness, freshness, documentation, provenance, licensing, and
reliability, all of which can affect their suitability for a target
task~\cite{chapman2020dataset,weissgerber2024provenance}. This creates a
practical need for quality-, provenance-, and trust-aware dataset search that
considers not only whether a dataset is relevant, but also whether it is
reliable, traceable, and appropriate for reuse.

Dataset quality assessment, provenance management, and trustworthiness
evaluation have been extensively studied in data management and scientific
data reuse~\cite{gong2023survey,miller2025guidelines}. However, these factors
are still commonly treated as curation or post-retrieval assessment rather
than being directly incorporated into dataset search and ranking. This gap
becomes increasingly important as automatically generated metadata,
descriptions, and other semantic evidence influence which datasets are
retrieved. Extending dataset search from relevance-based retrieval toward
reliable dataset selection raises several open challenges:

\begin{itemize}

    \item \textbf{Task-dependent Quality-aware Ranking.}
    Dataset quality is multi-dimensional and depends on the intended use:
    completeness, timeliness, coverage, bias, and documentation may matter
    differently for different tasks~\cite{chicco2025venus}. The challenge is
    therefore to jointly model relevance and task-dependent quality without
    reducing heterogeneous quality factors to a single fixed score.

    \item \textbf{Provenance-, Version-, and License-aware Search.}
    Datasets may be copied, derived, aggregated, and republished, making
    their original sources, modification histories, and usage rights
    difficult to trace. Missing or inconsistent provenance and license
    information can lead users to retrieve outdated, duplicated, or legally
    unsuitable datasets~\cite{longpre2024audit}. Future search systems need
    to preserve and exploit dataset lineage, version relationships, and
    usage constraints during ranking and result presentation.

    \item \textbf{Trustworthy LLM-mediated Dataset Search.}
    As LLMs increasingly generate metadata, summaries, pseudoqueries, and
    other searchable evidence, search systems may rank datasets according to
    information that is not fully grounded in the underlying data. Improving
    findability can therefore conflict with
    faithfulness~\cite{terrenzi2026faithful}. A key challenge is to make
    LLM-generated search evidence verifiable and provenance-aware so that
    improved retrieval does not come at the cost of user trust.

\end{itemize}

\subsection{Dataset Search for Data Preparation and Augmentation}

As reviewed in Sections~\ref{sec:ds} and
\ref{sec:llm_for_dataset_search}, existing work has begun to connect dataset
discovery with data augmentation through relationship search, utility-guided
selection, and agentic workflows~\cite{castelo2021auctus,
galhotra2023metam,tang2026lakeagents}. However, data preparation and
augmentation commonly operate on datasets already available to a pipeline,
while dataset discovery is often treated as a preceding
step~\cite{paton2023dataset,fernandes2023data}. This separation limits the
pipeline's ability to address missing attributes, records, or coverage
through additional data acquisition when such deficiencies become apparent
during processing. Moreover, retrieving individually relevant datasets may
produce redundant results that contribute little additional information to
the current data, since their value can depend on what has already been
selected or integrated. Result diversity has also been identified as an
important requirement for dataset search~\cite{hulsebos2024longer}. These
observations motivate tighter integration between dataset search and
preparation, in which search adapts to evolving data needs and candidates are
assessed according to their additional contribution to the assembled data.
Several challenges remain:

\begin{itemize}

    \item \textbf{Discovering Data for Evolving Needs.}
    A pipeline may identify missing information during execution, but
    translating these deficiencies into effective search queries is
    non-trivial. Useful data may also be distributed across heterogeneous
    repositories. Future systems should derive explicit data needs from
    intermediate results, identify appropriate sources, and discover
    additional datasets beyond the initial inputs.

    \item \textbf{Selecting Complementary Datasets and Operations.}
    Candidate utility depends on both the data already selected and the
    operations needed to combine them. Semantic dissimilarity alone does not
    guarantee useful complementarity, which may instead arise from additional
    attributes, records, or temporal coverage. Future methods should jointly
    select datasets and preparation operations, balancing relevance,
    additional coverage, expected task utility, and processing cost. This
    would help avoid redundant or suboptimal
    augmentation~\cite{turchenko2026matryoshka}.

    \item \textbf{Adapting Search through Preparation Feedback.}
    Failed joins, unsuitable intermediate results, or unmet task requirements
    may reveal that earlier search or integration decisions need
    revision~\cite{fan2026deepprep}. Future pipelines should use such
    feedback to refine search queries, reassess which data remain missing,
    and revise integration decisions. The desired complementarity may also
    change with the processing state, favoring broader alternatives during
    exploration and more targeted additions when specific deficiencies are
    identified.

\end{itemize}

\subsection{Comprehensive Benchmarks for Dataset Search}

As reviewed in Section~\ref{sec:benchmarks}, existing benchmarks provide
useful foundations for evaluating dataset search, but their coverage remains
uneven. Many focus on tabular collections, while benchmarks spanning multiple
dataset modalities often evaluate search primarily through textual
descriptions, metadata, or scientific literature. Search over non-tabular
contents, structural similarity, and cross-modal dataset retrieval remains
underrepresented, limiting systematic comparison of methods targeting these
settings. Recent benchmarks have also extended evaluation to explicit user
requirements~\cite{jin2026followtable,cui2025nlctables} and downstream
analytical workflows~\cite{wang2026lakeqa,lai2026kramabench}. However,
synthetic queries may simplify practical search needs, while relevance
judgments derived from observed dataset usage or candidate pooling may omit
other suitable datasets~\cite{viswanathan2023datafinder,deng2024lakebench}.
Building more comprehensive and reliable benchmarks therefore raises several
open challenges:

\begin{itemize}

    \item \textbf{Broader Coverage of Search Settings.}
    Future benchmarks should cover more dataset modalities and relevance
    criteria while distinguishing content-aware dataset search from search
    based only on textual descriptions or metadata. Shared collections
    supporting multiple user-input forms and relevance criteria would help
    examine how methods generalize across search settings.

    \item \textbf{Realistic User Needs and Reliable Judgments.}
    Benchmark construction should combine practical user needs with
    controlled queries that test specific requirements. Relevance assessment
    should consider multiple valid datasets and document how candidates are
    selected for annotation. Human validation of synthetic queries and
    model-generated labels, together with broader candidate pooling, could
    reduce construction bias and improve judgment coverage.

    \item \textbf{Separating Retrieval Quality from Downstream Utility.}
    For search embedded in analytical workflows, task success depends on
    retrieval as well as subsequent integration and reasoning. Future
    benchmarks should measure both retrieval and downstream outcomes while
    using controlled comparisons to isolate the contribution of dataset
    selection.

\end{itemize}

\section{Conclusions}\label{sec: Conclusion}

With the growing demand for {datasets} across diverse domains, open dataset
search has become an important infrastructure for data discovery and reuse.
{This survey has systematically reviewed recent advances in open dataset
search, covering its practical foundations, emerging search paradigms beyond
conventional keyword--metadata matching, the growing role of LLMs throughout
the search process, and representative benchmarks and evaluation practices.}
{Across} these studies, we observe a clear evolution {in how dataset relevance
is defined and assessed}. {Dataset search has expanded from metadata matching
toward richer notions of relevance based on dataset contents, inter-dataset
relationships, and task-specific suitability or utility.}
{Dataset representations and relevance estimation mechanisms have accordingly become
more fine-grained and structure-aware, while search is increasingly moving
from fixed query matching toward more adaptive, interactive, and
reasoning-based processes.}
{The growing integration of LLMs further extends
this evolution by supporting search evidence construction, dataset
representation and indexing, user input and search planning, and relevance
estimation.}
{At the same time, existing benchmarks reveal gaps in modality
coverage, realistic user needs, and the evaluation of retrieval and
downstream outcomes.}
{We further identify open issues and future directions
that may shape the next generation of dataset search systems.}
We hope this survey provides a structured view of {the evolving landscape} and supports
further advances in open dataset search.

\newpage
\end{sloppypar}